\documentclass[twocolumn,aps,prd,preprintnumbers,showpacs,showkeys,nofootinbib,
superscriptaddress,fleqn,floatfix,tightenlines]{revtex4-2}
\usepackage{amsmath,amsfonts,amssymb,amscd,amsxtra,amsthm}
\usepackage{graphicx}  
\usepackage{epstopdf}
\usepackage{dcolumn}  
\usepackage{bm}          
\usepackage{slashed}
\usepackage{cancel}
\usepackage{float} 
\usepackage{mathtools}
\usepackage{amsbsy}
\usepackage{amstext}

\usepackage[utf8]{inputenc} 
\usepackage{booktabs}
\usepackage[normalem]{ulem} 
\usepackage[dvipsnames]{xcolor} 
\usepackage{tabularx}
\usepackage{enumitem}  
\usepackage{array} 
\usepackage{slashed}
\usepackage{tikz}
\usepackage{float}
\usepackage{multirow}
\renewcommand\sout{\bgroup \color{red} \ULdepth=-.5ex \ULset}

\makeatletter

\begin{document}  
\sloppy
\preprint{INHA-NTG-03/2022}
\title{Medium modification of the nucleon mechanical properties:  
  Abel tomography case} 
\author{June-Young Kim}
\email[E-mail: ]{Jun-Young.Kim@ruhr-uni-bochum.de}
\affiliation{Ruhr-Universit\"at Bochum, Fakult\"at f\"ur Physik und
  Astronomie, Institut f\"ur Theoretische Physik II, D-44780 Bochum,
  Germany}   

\author{Ulugbek Yakhshiev}
\email[E-mail: ]{yakhshiev@inha.ac.kr}
\affiliation{Department of Physics, Inha University, Incheon 22212,
  Republic of Korea}  
\affiliation{Theoretical Physics Department, National University of
  Uzbekistan, Tashkent 100174, Uzbekistan}

\author{Hyun-Chul Kim}
\email[E-mail: ]{hchkim@inha.ac.kr}
\affiliation{Department of Physics, Inha University, Incheon 22212,
Republic of Korea}
\affiliation{School of Physics, Korea Institute for Advanced Study
  (KIAS), Seoul 02455, Republic of Korea}

\date{\today}
\begin{abstract}
We investigate how the gravitational form factors of the nucleon
undergo changes in nuclear matter, emphasizing the Abel transformation
from the three-dimensional (3D) Breit frame to the two-dimensional
(2D) light-front frame. Since the gravitational form factors reveal
the mechanical structure of the nucleon, we examine also the
medium modifications of the energy-momentum, pressure, and shear-force 
distributions. We scrutinize the stabilities of the
nucleon in nuclear matter. For this purpose, we employ the in-medium
modified SU(2) Skyrme model to study these mechanical quantities of
the nucleon, since it provides a simple but clear framework. In
this in-medium modified SU(2) Skyrme model, the modification of pionic
properties is performed by using low-energy pion-nucleus scattering  
data and the saturation properties of nuclear matter near the
normal nuclear matter density, $\rho_0=0.5m_\pi^3$.
The results reveal how the nucleon swells in nuclear
matter as the mass distribution of the nucleon is broadened in medium.  
We also show that the mean square radii corresponding to the mass
and angular momentum distributions increase in nuclear medium.  
This feature is kept both in 3D and 2D cases. We visualize how the
strong force fields inside the nucleon in the 2D plane are distributed
and illustrate how these forces undergo change in nuclear matter.   
\end{abstract}
\maketitle

\section{Introduction}
The gravitational form factors (GFFs) or the energy-momentum tensor
(EMT) form factors (FFs) of the nucleon provide essential 
information on the mass, spin, pressure, and shear-force
distributions. In particular, the force distributions inside a nucleon
shed light on how the nucleon acquires 
stability microscopically. While the GFFs were proposed many years
ago~\cite{Kobzarev:1962wt, Pagels:1966zza}, it is practically
impossible to measure them directly. However, the 
generalized parton distributions (GPDs) furnish a modern understanding
of the form factors. The electromagnetic form factors and GFFs are
identified as the first and second moments of the vector
GPDs~\cite{Muller:1994ses, Ji:1996ek, Radyushkin:1996nd} (see also
relevant reviews~\cite{Goeke:2001tz, Diehl:2003ny,
  Belitsky:2005qn}), respectively. It indicates that the GFFs of the
nucleon can be extracted from the experimental data on the vector 
GPDs~\cite{Burkert:2018bqq, Burkert:2021ith}. The GFFs reveal a unique
feature of the mechanical structure of the nucleon: Any hadron should  
satisfy the stability condition, also known as the von Laue
condition~\cite{Laue:1911lrk}, which arises from the conservation of
the EMT current (see, for example, a recent 
review~\cite{Polyakov:2018zvc}). It is a nontrivial condition and yet
any approach for describing the structure of the nucleon should
comply with it. 
 
The $D$-term form factor of the nucleon, which is one of the GFFs, is
associated with the pressure and shear-force distributions, which
carry information on the mechanism of the nucleon stabilization and
dictate a criterion for the stability condition of the
nucleon~\cite{Polyakov:2002yz,Polyakov:2018zvc, Lorce:2018egm}.   
These three-dimensional (3D) distributions of the pressure and
shear-force are often presented in the Breit frame
(BF)~\cite{Polyakov:2002yz, Goeke:2007fp, Polyakov:2018zvc}. 
On the other hand, the validity of the 3D densities for the
nucleon~\cite{Yennie:1957, Burkardt:2000za, Burkardt:2002hr,
  Belitsky:2005qn, Miller:2007uy, Miller:2010nz, Jaffe:2020ebz} has
been under question since the experiments on the proton 
structure performed by Hofstadter~\cite{Hofstadter:1956qs,
  Hofstadter:1956qs, Hofstadter:1957wk}. 
The criticisms have it that the 3D distributions are valid only for  
the nonrelativistic particles such as atoms and nuclei, the intrinsic 
sizes of which are much larger than the characteristic Compton  
wavelength ($\lambdabar=\hbar/mc$). However, when it comes to the
nucleon, one cannot define a localized state 
because the intrinsic size of the nucleon bears comparison with its
Compton wavelength. It implies that the nucleon is \emph{per se} a
relativistic particle~\cite{Belitsky:2005qn}, so that the
probabilistic interpretation of its 3D distributions in the BF is
tainted by relativistic corrections. 

Nevertheless, the BF distributions for the static EMT can be
interpreted as quasi-probabilistic densities from the phase-space
perspective\,\cite{Lorce:2017wkb, Lorce:2018zpf, Lorce:2018egm,
  Lorce:2020onh, Lorce:2021gxs}. It means that the BF distributions,
which are defined through the Wigner distributions in the quantum
phase space~\cite{Wigner:1932eb, Hillery:1983ms}, can be understood 
as quasi-probabilistic ones due to Heisenberg's uncertainty principle.
If one takes the infinite momentum frame (IMF), which indicates that
the nucleon is on the light-front (LF), the relativistic corrections
to the distributions are suppressed kinematically and a transversely
localized state for the nucleon can be defined, which enables one to
construct the two-dimensional (2D) transverse distributions on the
LF. They provide a strict probabilistic
interpretation~\cite{Burkardt:2002hr, Miller:2007uy, 
  Carlson:2007xd} and are not Lorentz contracted since the initial
and final states of the nucleon lie on the mass shell. Recently, it
was shown in Ref.~\cite{Lorce:2020onh, Kim:2021kum, Lorce:2022jyi}
that the BF charge distributions could be interpolated to the LF
ones. In Ref.~\cite{Panteleeva:2021iip}, this interpolation was
applied to the EMT force distributions as well as the electromagnetic 
ones~\cite{Kim:2021jjf, Kim:2022bia}. The Abel transform~\cite{Abel},
which has been used in the computerized medical
tomography~\cite{Natterer:2001}, brings the 3D BF distributions to the
2D LF ones in the transverse plane. It was already utilized in the
deeply virtual Compton scattering~\cite{Polyakov:2007rv,
  Moiseeva:2008qd}. As pointed out in Ref.~\cite{Panteleeva:2021iip},
it is crucial to scrutinize how the EMT force distributions in the BF
are related to those in the IMF. 

In Ref.~\cite{Freese:2021mzg} the Abel transformation from the the 3D
BF distributions to the 2D IMF ones was criticized. That in
Ref.~\cite{Freese:2021mzg} was defined by integrating a spherical
symmetric 3D distribution merely over $z$-axis. It means that the 
integral projects the 3D BF distribution onto the 2D BF one without
any relativistic effects. Thus, the Abel transformation defined in
Ref.~\cite{Freese:2021mzg} is restricted within the BF, so that   
their Abel images are different from those defined in
Refs.~\cite{Kim:2021jjf, Panteleeva:2021iip}. As far as the EM and EMT
transverse densities of the nucleon are concerned, we can project the 
3D distributions onto the 2D transverse densities in
the LF. Contrary to the criticism raised in
Ref.~\cite{Freese:2021mzg}, the 2D IMF transverse charge densities of
the nucleon~\cite{Miller:2007uy} were successfully reproduced by
Abel-transforming the nonrelativistic 3D charge and magnetization
distributions in Ref.~\cite{Kim:2021jjf}. Thus, the criticism of 
Ref.~\cite{Freese:2021mzg} was not applied to the present
definition of the Abel transform. We will show in the current work
that the EMT transverse densities in the LF are indeed derived by the
Abel transformation of the EMT distributions in the 3D BF as done in
Ref.~\cite{Panteleeva:2021iip}.   

It is also of great interest to examine how the mechanical properties
of the nucleon undergo modification in nuclear matter. In fact, the
GFFs of the nucleon in nuclear medium were already studied in
Ref.~\cite{Kim:2012ts}. In the present work, we aim at providing the
genuine 2D EMT distributions of the nucleon in nuclear matter in the
IMF. To investigate them, we employ the in-medium modified Skyrme
model~\cite{Rakhimov:1996vq, Rakhimov:1998hu,Kim:2012ts}. The model is
known to be one of the simplest ones for describing the lowest-lying
baryons based on the $1/N_{c}$ expansion. In the large $N_{c}$ limit,
a baryon arises as a chiral soliton with effective mesonic degrees of 
freedom\,\cite{Adkins:1983ya} on account of the suppression of the
meson fluctuation. In addition, the model is related to
the essential properties of the QCD, such as chiral symmetry and its
spontaneous breaking. 

Qualitatively, the model quite satisfactorily
describes the important properties of the nucleon at low-energy
regime. The merit of the model is that it can be easily applied to
study the numerous baryon properties and has even described well the
general properties of the nucleon GFFs. This model was already 
extended to consider the nucleon properties in nuclear matter by
modifying the properties of mesons in medium~\cite{Rakhimov:1996vq,
  Rakhimov:1998hu}. In Ref.~\cite{Yakhshiev:2010kf}, the model has
been further elaborated to consider the nuclear matter properties at
saturation density where the stabilizing term was refined in nuclear
medium. In the present work, we will take Ref.~\cite{Yakhshiev:2010kf}
as our model framework to revisit the GFFs and the 3D distributions of
the nucleon in nuclear matter and study the modification of the 2D
images of the nucleon pressure and shear-force distributions. 

The present work is organized as follows: In Section II, we formulate
the 2D and 3D EMT distributions of the nucleon in general. In Section
III, we show how to compute the GFFs of the nucleon in nuclear
matter. In Section IV, we present the results and discuss them,
focusing on how the strong force fields are distributed inside a
nucleon in nuclear matter. The final section summarizes the present
work draws conclusions.

\section{2D and 3D EMT distributions of the nucleon}
We first recapitulate the GFFs of the 
nucleon~\cite{Kobzarev:1962wt,Pagels:1966zza}. The nucleon matrix
element of symmetric EMT operator $\hat{T}^{\mu\nu}(0)$ between the 
nucleon states with initial(final) momentum $p (p')$ and helicity 
$\lambda (\lambda')$ can be parametrized in terms of the three form
factors  $A(t)$, $J(t)$, and $D(t)$ as follows 
\begin{align}
\langle p',\lambda' &| \hat{T}^{\mu\nu}(0) | p, \lambda \rangle =
  \bar{N}_{\lambda'}(p')\bigg{[} A(t) \frac{P^{\mu}P^{\nu}}{m} \cr
& + J(t) \frac{i P^{\{\mu}\sigma^{\nu\}\alpha} \Delta_{\alpha}}{m} \cr
& +\frac{D(t)}{4m} (\Delta^{\mu}\Delta^{\nu}-g^{\mu\nu} \Delta^{2})
  \bigg{]} N_{\lambda}(p)\,,
  \label{eq:EMTff}
\end{align}
where ${N}_{\lambda}(p)$ is the Dirac spinor which is normalized as
$\bar{N}_{\lambda'}(p)N_{\lambda}(p) = 2m \delta_{\lambda'\lambda}$
and we introduced kinematic variables $P^{\mu}=(p^{\mu}+p'^{\mu})/2$, 
$\Delta^{\mu}=p'^{\mu}-p^{\mu}$, and $\Delta^{2} = t$. The
parenthesis $a^{\{\mu}b^{\nu\}}=(a^{\mu}b^{\nu}+a^{\nu}b^{\mu})/2$
stands for the symmetrization operator and $m$ is the
nucleon mass. We also use the covariant normalization $\langle
p', \lambda'| p, \lambda \rangle = 2p^{0} (2\pi)^{3}\delta_{\lambda'
  \lambda} \delta^{(3)}(\bm{p}'-\bm{p})$ for the one-particle
states. Three different form factors, $A(t)$, $J(t)$ and $D(t)$,
provide information on the mass, spin, and the mechanical
properties of the nucleon, respectively.  
The mass form factor is given as the following linear combination 
\begin{align}
M(t) = A(t) - \frac{t}{4m^2} \bigg{(}A(t)
  -2J(t) +D(t)\bigg{)}. 
\label{eq:mass}
\end{align}
In the Wigner sense, the mass $\varepsilon(r)$, angular momentum
$\rho_{J}(r)$, pressure $p(r)$ and shear-force $s(r)$ distributions in 
the BF are obtained as 
\begin{align}
\varepsilon(r) &=  m \tilde{M}(r), \cr 
\rho_{J}(r)&= -\frac{1}{3}r \frac{d}{dr} \tilde{J}(r), \cr s(r)& =
  -\frac{1}{4m} r \frac{d}{dr} \frac{1}{r} \frac{d}{dr} \tilde{D}(r),\cr
   p(r)&=
  \frac{1}{6m}\frac{1}{r^{2}}\frac{d}{dr}r^{2}\frac{d}{dr}\tilde{D}(r), 
\label{eq:3DFF}
\end{align}
by using the generic 3D inverse Fourier transform 
\begin{align}
\tilde{F}(r) = \int \frac{d^{3}
  \bm{\Delta}}{(2\pi)^{3}}e^{-i\bm{\Delta} \cdot \bm{r}}
  F(-\bm{\Delta}^{2}),
\end{align}
of the GFFs, where $F=A, M, J, D$ are defined
in terms of the multipole expansion~\cite{Lorce:2018egm, 
  Polyakov:2018zvc, Polyakov:2018rew, Kim:2020lrs}. 

However, the physical meaning of 3D distributions is marred by
ambiguity because of the relativistic corrections. This ambiguity can
be remedied by considering the 2D transverse distributions in the
IMF~\cite{Burkardt:2000za, Burkardt:2002hr, Belitsky:2005qn,
  Miller:2007uy, Miller:2010nz, Jaffe:2020ebz}.  
The EMT distributions in the IMF have been already extensively studied
in numerous Refs.~\cite{Lorce:2018egm, 
  Freese:2021czn, Lorce:2017wkb, Schweitzer:2019kkd, Lorce:2018egm,
  Panteleeva:2021iip, Epelbaum:2022fjc, Lorce:2022cle}. 
The corresponding EMT 2D distributions for the energy
$\varepsilon^{(2D)}(x_{\perp})$, angular momentum
$\rho_{J}^{(2D)}(x_{\perp})$, pressure $p^{(2D)}(x_{\perp})$ and shear
force $s^{(2D)}(x_{\perp})$, which have the following 
forms 
\begin{align}
& \varepsilon^{(2D)}(x_{\perp}) =  P^{+} \tilde{A}(x_{\perp}), \cr
&\rho^{(2D)}_{J}(x_{\perp})= -\frac{1}{2}x_{\perp} \frac{d}{dx_{\perp}}
  \tilde{J}(x_{\perp}), \ \cr
&s^{(2D)}(x_{\perp}) = -\frac{1}{4P^{+}}
   x_{\perp}\frac{1}{dx_{\perp}} \frac{1}{x_{\perp}}
   \frac{d}{dx_{\perp}}  \tilde{D}(x_{\perp}), \cr
   & p^{(2D)}(x_{\perp})=
   \frac{1}{8P^{+}}\frac{1}{x_{\perp}}
   \frac{d}{dx_{\perp}}x_{\perp}\frac{d}{dx_{\perp}}\tilde{D}(x_{\perp}),  
 \label{eq:2DFF}
\end{align}
were obtained by using the generic 2D inverse Fourier
transform 
\begin{align}
\tilde{F}(x_{\perp}) = \int \frac{d^{2}
  \bm{\Delta}_{\perp}}{(2\pi)^{2}}e^{-i\bm{\Delta}_{\perp} \cdot
  \bm{x_{\perp}}} F(-\bm{\Delta}^{2}_{\perp}). 
\end{align}
of the corresponding GFFs. In the expressions above $\bm{x}_{\perp}$
and $\bm{\Delta}_{\perp}$ denote the position and momentum vectors in
the 2D plane transverse to the moving direction of the nucleon,
respectively. Consequently,  the mass, shear force, and pressure 
distributions are redefined by multiplying, for
convenience, the Lorentz factors~\cite{Panteleeva:2021iip} as follows
\begin{align}
\mathcal{E}(x_\perp)&= \frac{m}{P^{+}}\varepsilon^{(2D)}(x_\perp), \cr
  \mathcal{S}(x_\perp)&= \frac{P^{+}}{2m}s^{(2D)}(x_\perp),  \cr
  \mathcal{P}(x_\perp)&=\frac{P^{+}}{2m}p^{(2D)}(x_\perp),
\label{eq:2DFFcov}
\end{align}
where $P^{+}$ is the light-cone momentum. 

The BF and LF distributions given respectively in
Eq.~\eqref{eq:3DFF} and Eq.~\eqref{eq:2DFFcov} should be related to
each other and the corresponding relations were derived in 
Refs.~\cite{Panteleeva:2021iip, Kim:2021jjf} 
by the Abel transform
\begin{align}
\left(1 -
  \frac{\partial^{2}_{(2D)}}{4m^{2}}\right)\mathcal{E}(x_{\perp})& =
 2  \int^{\infty}_{x_{\perp}}
   \frac{rdr}{\sqrt{r^{2}-x^{2}_{\perp}}}
   \bigg[
   \varepsilon(r)\cr
  & +
   \frac{3}{2}p(r)
   +
   \frac{3}{2mr^{2}}\frac{d}{dr}r
   \rho_{J}(r)\bigg], \cr
  \rho^{(2D)}_{J}(x_{\perp})&= 3 \int^{\infty}_{x_{\perp}}
  \frac{\rho_{J}(r)}{r}
  \frac{x^{2}_{\perp}dr}{\sqrt{r^{2}-x^{2}_{\perp}}}, \cr
\mathcal{S}(x_{\perp}) &=\int^{\infty}_{x_{\perp}} \frac{s(r)}{r}
  \frac{ x^{2}_{\perp}dr}{\sqrt{r^{2}-x^{2}_{\perp}}}, 
  \cr  
  \frac{1}{2}\mathcal{S}(x_{\perp})+\mathcal{P}(x_{\perp})&
  =\frac{1}{2}\int^{\infty}_{x_{\perp}}
  \left(\frac{2}{3}s(r)+p(r)\right) \cr
  &\times \frac{ r
  dr}{\sqrt{r^{2}-x^{2}_{\perp}}}. 
\label{eq:Abel_PS}
\end{align} 
In the large $N_{c}$ limit, it is possible to derive the
reduced expression for the 2D energy distribution 
\begin{align}
\mathcal{E}(x_{\perp}) =
 2  \int^{\infty}_{x_{\perp}}
   \frac{rdr}{\sqrt{r^{2}-x^{2}_{\perp}}}
   \left[
   \varepsilon(r)
   +
   \frac{3}{2}p(r)\right]. 
\label{eq:Abel_PS_red}
\end{align} 
Integrating 2D and 3D EMT distributions over $\bm{x}_{\perp}$ and
$\bm{r}$, respectively, one can get the mass and spin of the nucleon 
\begin{align}
mA(0)&=\int d^{2}x_{\perp} \, \mathcal{E}(x_{\perp})=\int d^{3}r \,
       \varepsilon(r), \cr 
 J(0)&= \int 
  d^{2}x_{\perp} \, \rho_{J}^{(2D)}(x_{\perp}) = \int
  d^{3}r \, \rho_{J}(r),  
\label{eq:normal_2DEMT}
\end{align}
with the properly normalized form factors, $A(0)=1$ and
$J(0)=1/2$. As in the case of the radii for the 3D
distributions~\cite{Kim:2020nug}, we can define 2D radii for the mass   
and angular momentum 
\begin{align}
\langle  x^{2}_{\perp} \rangle_{\mathcal{E}} &=\frac{1}{m}\int
  d^{2}x_{\perp} x^{2}_{\perp} \mathcal{E}(x_{\perp})
  \cr
  &=\frac{2}{3}\langle r^{2} \rangle_{\varepsilon}
  +\frac{D(0)}{m^{2}}, \cr 
 \langle x^{2}_{\perp}
\rangle_{J}&=2\int d^{2}x_{\perp} x^{2}_{\perp}
  \rho^{(2D)}_{J}(x_{\perp})= \frac{4}{5} \langle r^{2} \rangle_{J},
\label{eq:radius}  
\end{align}
respectively\,\cite{Panteleeva:2021iip}.

The conservation of the EMT current also provides the 2D stability 
condition for the nucleon as in the 3D case,  which yields the
following 2D stability equation 
\begin{align}
 \mathcal{P}'(x_{\perp})+
  \frac{\mathcal{S}(x_{\perp})}{x_{\perp}} + 
  \frac{1}{2}\mathcal{S}'(x_{\perp})=0. 
\label{eq:diff_eq2}
\end{align}
This 2D stability condition can be considered to be equivalent to the
3D expression 
\begin{align}
p'(r)+ \frac{2s(r)}{r} +  \frac{2}{3}s'(r)=0. 
\label{eq:diff_eq3}
\end{align}
One can see that the shear-force and
pressure distributions are related to each other. Using
Eqs.~\eqref{eq:diff_eq2} and~\eqref{eq:diff_eq3}, we obtain the global
stability condition or the von Laue condition
\begin{align}
\int d^{3}r &\, p(r)=0 \iff
\int d^{2}{x}_{\perp} \mathcal{P}(x_{\perp})=0, 
\label{eq:stabp}
\\
 \int^{\infty}_{0} dr \, r&  \left[p(r)-\frac{1}{3}s(r)\right]=0 \iff
 \cr
 &\int^{\infty}_{0} d{x}_{\perp} \left[\mathcal{P}(x_{\perp})- 
  \frac{1}{2}\mathcal{S}(x_{\perp})\right]=0,  
\label{eq:stab}
\end{align}
which are the necessary and sufficient conditions for the stability of
the nucleon. 
The integrand in the last equation in Eq.~\eqref{eq:Abel_PS} gives the 
3D local stability condition, which leads to the 2D local stability 
condition in the LF~\cite{Panteleeva:2021iip} 
\begin{align}
\frac{2}{3}s(r)+p(r) >0 \iff
\frac{1}{2}\mathcal{S}(x_{\perp})+\mathcal{P}(x_{\perp}) >0. 
\label{eq:loc_stab}
\end{align}
The positivity is kept intact by the Abel transformation.
This implies that though the 3D distributions have only
quasi-probabilistic meaning, they still provide intuitive features for
the stability conditions. It also relates the 3D mechanical radius to
the 2D one  
\begin{align}
\langle x^{2}_{\perp} \rangle_{\mathrm{mech}} &= \frac{\int
  d^{2}x_{\perp} x^{2}_{\perp} \left(\frac{1}{2}\mathcal{S}(x_{\perp})
  +\mathcal{P}(x_{\perp})\right)}{\int d^{2}x_{\perp}
  \left(\frac{1}{2}\mathcal{S}(x_{\perp})
  +\mathcal{P}(x_{\perp})\right)} \cr
  &= \frac{4D(0)}{\int^{0}_{-\infty}dt
  D(t)} = \frac{2}{3}\langle r^{2} \rangle_{\mathrm{mech}}. 
\label{eq:mech2d}
\end{align}

To understand the physics of the pressure and shear-force
distributions more in detail, it is instructive to introduce the
notion of the normal and tangential force fields 
that are just the eigenvalues of the stress tensor contracted with the
radial $\bm{e}_{r}$ and tangential $ \bm{e}_{\phi}$ unit vectors,  
respectively. The 3D and the 2D force fields in the BF and LF
are expressed in terms of the pressure and shear-force distributions
as follows: 
\begin{align}
&F_{n}(r)=4\pi
  r^{2}\left[\frac{2}{3}s(r)
  +p(r)\right], \cr
&  \ F_{t}(r)= 4\pi
  r^{2}\left[-\frac{1}{3}s(r)
  +p(r)\right],   \cr
&F_{n}^{(2D)}(x_{\perp})=2\pi
  x_{\perp}\left[\frac{1}{2}\mathcal{S}(x_{\perp})
  +\mathcal{P}(x_{\perp})\right], \cr
 &F_{t}^{(2D)}(x_{\perp})= 2\pi
  x_{\perp}\left[-\frac{1}{2}\mathcal{S}(x_{\perp})
  +\mathcal{P}(x_{\perp})\right].  
\label{eq:2Dforce}
\end{align}
Equation~\eqref{eq:2Dforce} implies that $F_n(r)$ and
$F_n^{(2D)}(x_\perp)$ should be positive because of the local
stability condition in Eq.~\eqref{eq:loc_stab}, whereas $F_t(r)$ and
$F_t^{(2D)}(x_\perp)$ should have at least one nodal point to ensure
the von Laue conditions, which are shown in Eq.~\eqref{eq:stab}. We
will see these characteristics of the force fields later in detail. 

The value of the $D$-term is determined by integrating over the 3D and
2D pressure or shear-force distributions as  
\begin{align}
D(0) &= - \frac{4m}{15} \int d^{3}r \, r^{2} s(r)
= - m \int d^{2}x_{\perp} x^{2}_{\perp} \mathcal{S}(x_{\perp}) \cr
     &=   m \int d^{3}r \, r^{2} p(r) =4
  m \int d^{2}x_{\perp} x^{2}_{\perp} \mathcal{P}(x_{\perp}). 
\label{eq:dterm}
\end{align}
As will be shown soon, both the 2D and 3D shear-force distributions
are positive in the overall ranges of $x_\perp$ and $r$,
respectively. So, the $D(0)$ should be always negative, which is also
deeply related to the stability conditions.   
Equation~\eqref{eq:dterm} also shows the equivalence between the 2D
and 3D pressure and shear-force distributions. It implies that while
3D distributions are quasi-probabilistic, the 2D distributions
provide the proper probabilistic meaning.

\section{Grativational form factors of the nucleon in nuclear
  matter \label{sec:3}} 
To compute the GFFs of the nucleon in nuclear matter, we will employ 
the in-medium modified chiral soliton model. We start with the
in-medium modified chiral Lagrangian\footnote{Hereafter, a superscript
  ``${}^*$" indicates an modified in-medium
    quantity. We consider the in-medium modifications of the
    quantities discussed so far with the superscript
    included.}~\cite{Yakhshiev:2010kf,Kim:2012ts}  
\begin{align}
\mathcal{L}^{*}&=
    \mathcal{L}^{*}_2+\mathcal{L}^{*}_4+\mathcal{L}^{*}_{\rm  m}\cr 
&=\frac{F^{2}_{\pi}}{16}\mathrm{Tr}\left[\partial_{0}
  U\partial^{0}U^{\dagger} \right] 
+\alpha_{p} \frac{F^{2}_{\pi}}{16}\mathrm{Tr}\left[\partial_{i}
  U\partial^{i}U^{\dagger} \right]\cr 
& + \frac{1}{32e^{2}\gamma}\mathrm{Tr}
\left[(\partial_{\mu} U)U^{\dagger},
(\partial_{\nu} U)U^{\dagger} \right]^{2} \cr
&+ \alpha_{s}\frac{m^{2}_{\pi}F^{2}_{\pi}}{8}
\mathrm{Tr}\left[U-1\right],
\end{align}
where $U=\mathrm{exp}[\hat{r}^{i}\tau^{i} P(r)]$ is the SU(2) chiral
field with the profile function $P(r)$ for the pion fields,
$\tau^i\,(i=1,2,3)$ are the Pauli matrices, and
$\hat{r}^{i}={r^{i}}/{|\bm{r}|}$ is the radial component of the unit 
vector in space. $\mathrm{Tr}$ stands for the trace running over the
SU(2) isospin space. The input parameters of the soliton model are
given by the pion decay constant $F_\pi=108.78$\,MeV, a dimensionless 
parameter $e=4.854$, and the pion mass $m_\pi=135$\,MeV. 

We incorporate the medium modifications by introducing the  
density-dependent medium functions  
\begin{align}
\alpha_{p}(\rho) &= 1 -\chi_{p}(\rho), \, \, \, \chi_{p}(\rho) =
                   \frac{4\pi c_{0} \rho}{ \eta + 4\pi c_{0}g' \rho},
                   \, \, \cr  
\eta&= 1+ \frac{m_{\pi}}{m}, \cr
\alpha_{s}(\rho) &= 1 +\frac{\chi_{s}(\rho)}{m^{2}_{\pi}}, \, \, \,
                   \chi_{s}(\rho) = -4\pi \eta b_{0} \rho, \cr 
\gamma(\rho)
  &=\mathrm{exp}\left[-\frac{\gamma_{\mathrm{num}}\rho}{
1+\gamma_{\mathrm{den}}\rho}\right], 
\end{align}
which associate the model with the low energy pion-nucleus scattering
data and properties of nuclear matter at the normal nuclear matter 
density or the saturation point $\rho_0$~\cite{Yakhshiev:2010kf}.
The parameters in the medium functions have been already fixed by
\begin{align}
b_{0}& = -0.024\,m_{\pi}^{-1}, \ \ \ c_{0} = 0.09\,m_{\pi}^{-3}, \ \ \
 g'=0.7, \cr \gamma_{\mathrm{num}}&=0.797\,m_{\pi}^{-3}, \ \ \ 
\gamma_{\mathrm{den}}=0.496\,m_{\pi}^{-3},
\end{align}
where $b_0$ and $c_0$ are s- and p-wave pion-nucleon scattering
lengths and volumes, $g'$ is correlation parameter, and $\gamma_{\rm 
num}$ and $\gamma_{\rm den}$ are fitted in such a way that nuclear
matter properties at the normal nuclear matter density $\rho_0=0.5
m^{3}_{\pi}$ are reproduced correctly. For details we refer to
Ref.~\cite{Yakhshiev:2010kf}. 

Since we consider homogeneous nuclear matter, where the nuclear density
is kept constant, we treat the modified chiral Lagrangian by
introducing the renormalized effective constants:
\begin{align}
F^{*}_{\pi,t}&=F_{\pi,t}=F_{\pi}, \ \ \ F^{*}_{\pi,s}
               =\alpha^{1/2}_{p}F_{\pi}, \cr  
e^{*}&=\gamma^{1/2}e, \ \ \
       m^{*}_{\pi}=(\alpha_{s}/\alpha_{p})^{1/2}m_{\pi}, 
\end{align}
which shows that all expressions can be expressed in terms of the
renormalized model parameters. For example, the classical soliton mass
functional  in nuclear matter takes the following form 
\begin{align}
M^{*}_{\mathrm{sol}}[P]&=4\pi \int^{\infty}_{0}dr \,r^{2} \cr
&\times
\left[\frac{F^{*2}_{\pi,s}}{8}\left(\frac{2\mathrm{sin}^{2}P(r)}{r^{2}}
  + P'(r)^{2}\right)\right. \cr 
&+  \frac{\mathrm{sin}^{2}P(r)}{2e^{*2}r^{2}}
\left(\frac{\mathrm{sin}^{2}P(r)}{r^{2}}
  + 2P'(r)^{2}\right) \cr 
&\left.+ \frac{m^{*2}_{\pi}F^{*2}_{\pi,s}}{4}
\left(1-\mathrm{cos}P(r)\right)\right],
\label{Mstat}
\end{align}
where $P'(r)$ denotes the derivative with respect to the 
variable $r$. Minimizing the soliton mass, we obtain the nonlinear
differential equation for the profile function $P(r)$ and the solution
for the baryon number $B=1$ is derived by imposing the boundary
conditions, $P(0)= \pi$ and $P(\infty) = 0$. Then making a zero-mode
quantization or rotating the classical soliton slowly, we get the
time-dependent SU(2) soliton $U(\bm{r})\rightarrow A(t)
U(\bm{r})A^\dagger(t)$ with the collective
Hamiltonian~\cite{Adkins:1983ya} 
\begin{align}
H^{*}= M^{*}_{\mathrm{sol}} + \frac{\hat{\bm{J}}^{2}}{2I^{*}},
\end{align}
where $\hat{\bm{J}}$ is angular momentum operator and  
\begin{align}
I^{*}&= \frac{2\pi}{3} \int_0^\infty dr \, r^{2} \mathrm{sin}^{2}P(r) \cr
&\times\left[ F_{\pi}^{2} + \frac{4P'(r)^{2}}{e^{*2}} 
+ \frac{4\mathrm{sin}^{2}P(r)}{e^{*2}r^{2}} \right]
\end{align}
is  the moment of inertia of the rotating soliton. We derive 
the EMT from the time-dependent Lagrangian for the rotating soliton   
\begin{align}
T^{\mu \nu *} = \frac{\partial \mathcal{L}^{*}}{\partial 
(\partial_{\mu} \phi_{a})} \partial^{\nu} \phi_{a} - g^{\mu \nu} \mathcal{L}^{*},
\end{align}
where $\phi_{a}$ is a time-dependent pionic field with $U(t,r) =
\phi_{0}+ i \bm{\tau}\cdot\bm{\phi}$ that satisfied
the unitarity condition $\phi^{2}_{0} + \bm{\phi}^{2}=1$.  The
components of the EMT are then written as 
\begin{align}
T^{00*}
&= \delta_{\sigma'\sigma}\left[\frac{F^{*2}_{\pi,s}}{8}
\left(\frac{2\mathrm{sin}^{2}P(r)}{r^{2}} + P'(r)^{2}\right) \right.\cr
&+ \frac{\mathrm{sin}^{2}P(r)}{2e^{*2}r^{2}}
\left(\frac{\mathrm{sin}^{2}P(r)}{r^{2}} + 2P'(r)^{2}\right) \cr
&\left.+ \frac{m^{*2}_{\pi}F^{*2}_{\pi,s}}{4}
\left(1-\mathrm{cos}P(r)\right)\right],\cr
T^{ij*}
&= \hat{r}^{i}\hat{r}^{j}\delta_{\sigma'\sigma}\left[
\left(\frac{F^{*2}_{\pi,s}}{4} + 
\frac{\mathrm{sin}^{2}P(r)}{e^{*2}r^{2}}\right)\right.\cr
&\left.\times\left(P'(r)^{2} - 
\frac{\mathrm{sin}^{2}P(r)}{r^{2}} \right)\right] \cr
&+ \delta^{ij}\delta_{\sigma'\sigma}
\left[-\frac{F_{\pi,s}^{*2}}{8}P'(r)^{2} + 
\frac{\mathrm{sin}^{4}P(r)}{2e^{*2}r^{4}} \right.\cr
&\left.- \frac{m^{*2}_{\pi}F^{*2}_{\pi,s}}{4}
\left(1-\mathrm{cos}P(r)\right)\right],\cr
T^{0k*} 
&= (\hat{\bm{J}}\times\hat{\bm{r}})^{k}_{\sigma'\sigma} 
\frac{\mathrm{sin}^{2}P(r)}{4I^{*}r}\cr
&\times \left[F_{\pi}^{2} +\frac{4\mathrm{sin}^{2}P(r)}{e^{*2}r^{2}} 
+ \frac{4P'(r)^{2}}{e^{*2}}\right].
\label{eq:EMT_leading}
\end{align}
The time component $T^{00*}$ yields the leading-order 
term for the energy, i.e., for the static mass distribution given by
the integrand in Eq\,(\ref{Mstat}). The mixed space-time  
components $T^{0k*}$ give the angular momentum density whereas the
spatial ones $T^{ij*}$ furnish the pressure and shear-force densities.

We can extract the energy, spin, pressure, and shear-force
distributions directly from Eq.~\eqref{eq:EMT_leading}: 
\begin{align}
\varepsilon^{*}(r) &= \left[\frac{F^{*2}_{\pi,s}}{8}
\left(\frac{2\mathrm{sin}^{2}P(r)}{r^{2}} + P'(r)^{2}\right) \right.\cr
&+ \frac{\mathrm{sin}^{2}P(r)}{2e^{*2}r^{2}}
\left(\frac{\mathrm{sin}^{2}P(r)}{r^{2}} + 2P'(r)^{2}\right)\cr
&\left.+ \frac{m^{*2}_{\pi}F^{*2}_{\pi,s}}{4}
\left(1-\mathrm{cos}P(r)\right)\right], \cr
\rho^{*}_{J}(r)&= \frac{\mathrm{sin}^{2}P(r)}{12I^{*}} 
\left[F_{\pi}^{2} +\frac{4\mathrm{sin}^{2}P(r)}{e^{*2}r^{2}} 
+ \frac{4P'(r)^{2}}{e^{*2}}\right], \cr
p^{*}(r) &= -\frac{F^{*2}_{\pi,s}}{24}
\left(\frac{2\mathrm{sin}^{2}P(r)}{r^{2}} + P'(r)^{2}\right)\cr
& + \frac{\mathrm{sin}^{2}P(r)}{6e^{*2}r^{2}}
\left(\frac{\mathrm{sin}^{2}P(r)}{r^{2}} + 2P'(r)^{2}\right) \cr
&- \frac{m^{*2}_{\pi}F^{*2}_{\pi,s}}{4}
\left(1-\mathrm{cos}P(r)\right), \cr
s^{*}(r) &= \left(\frac{F^{*2}_{\pi,s}}{4}
+\frac{\mathrm{sin}^{2}P(r)}{e^{*2}r^{2}}\right)\cr
&\times\left(P'(r)^{2}-\frac{\mathrm{sin}^{2}P(r)}{r^{2}} \right).
\label{eq:distributions}
\end{align}
We arrive at the final expressions for the GFFs in Eq.(\ref{eq:EMTff})
given in the large $N_c$ limit as
\begin{align}
A^*(t)&-\frac{t}{4m^{*2}}\,D^*(t)
        = \frac{1}{m^*}\int\mathrm{d}^3
        \bm{r}\cr
        &\times \varepsilon^*(r)\;j_0(r\sqrt{-t})\,,
        \label{Eq:M2-d1-model-comp}\\
        D^*(t)
        &= 6 m^*\int\mathrm{d}^3 \bm{r} \;p^*(r)
        \;\frac{j_0(r\sqrt{-t})}{t} \,,
        \label{Eq:d1-model-comp}\\
        J^*(t)
        &= 3
    \int\mathrm{d}^3\bm{r}\;\rho_J^*(r)\;\frac{j_1(r\sqrt{-t})}{r\sqrt{-t}}\;, 
    \label{Eq:J-model-comp}
\end{align}
where $j_0(z)$ and $j_1(z)$ represent the spherical Bessel functions
of order 0 and 1, respectively. At the zero momentum transfer $t=0$,
$A^*(0)$ and $J^*(0)$ are normalized to be 
\begin{align}
  \label{eq:norm}
A^*(0)&= \frac{1}{m^*}
\int\mathrm{d}^3 \bm{r}\;\varepsilon^*(r) = 1 \,,\cr
 J^*(0) &=\int\mathrm{d}^3\bm{r}\;\rho_J^*(r)=\frac12\,.
\end{align}
These normalizations were proven in Ref.~\cite{Cebulla:2007ei} for a
Skyrmion in free space and they hold also in nuclear
matter~\cite{Kim:2012ts}. 

\section{Results and discussion \label{sec:4}}

\begin{figure*}[thp]
\begin{center}
\includegraphics[scale=0.265]{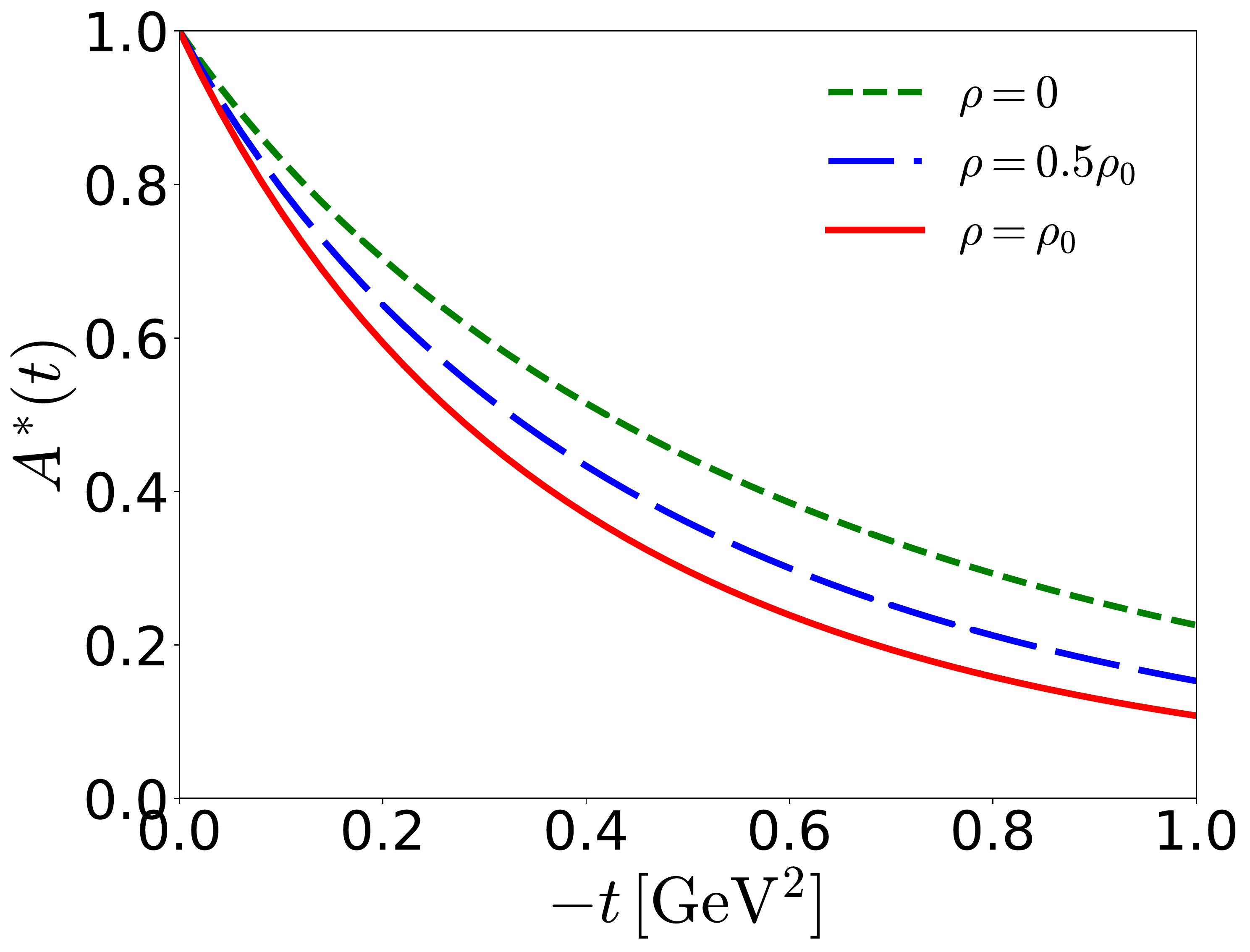}
\hspace{0.75cm}
\includegraphics[scale=0.265]{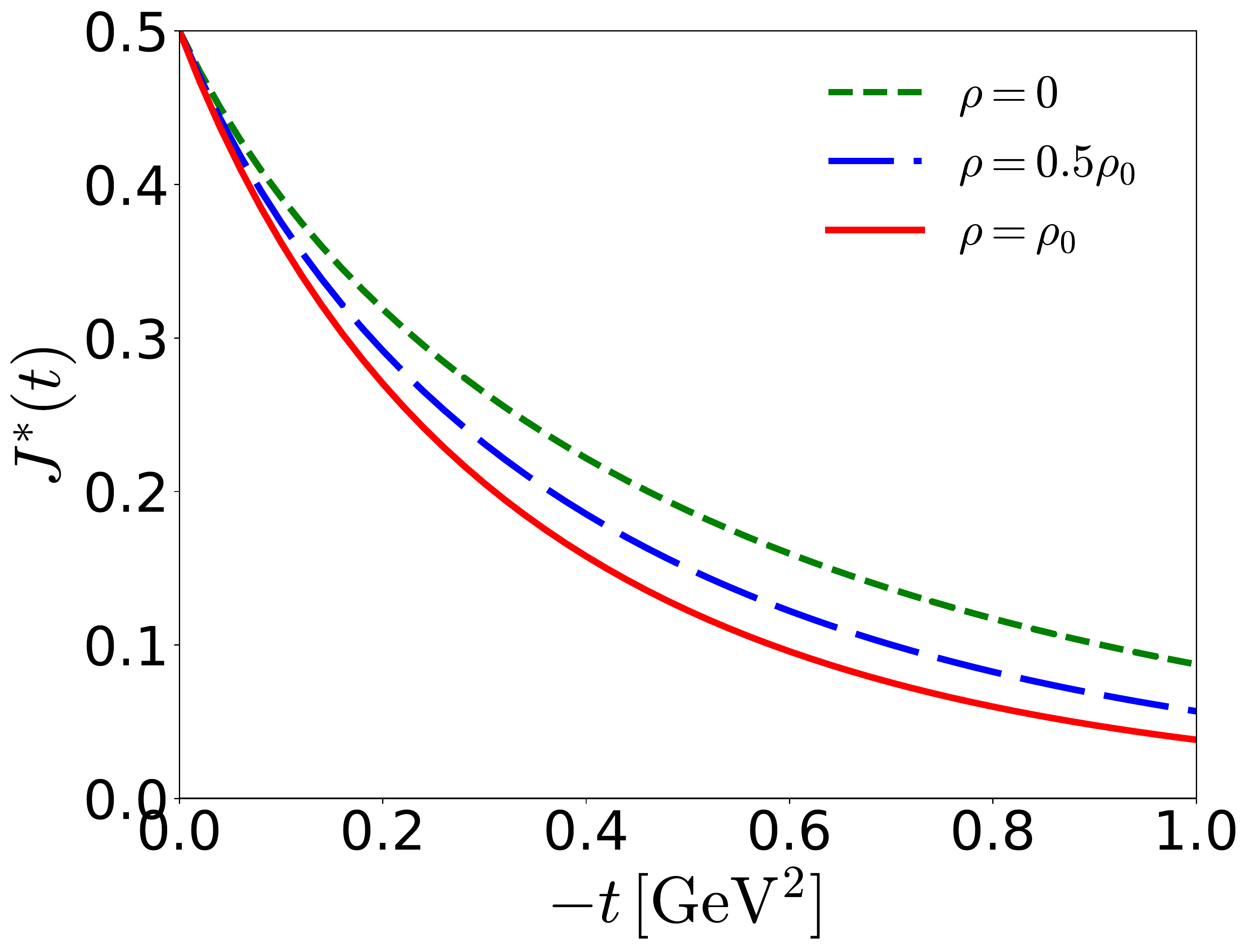}
\includegraphics[scale=0.265]{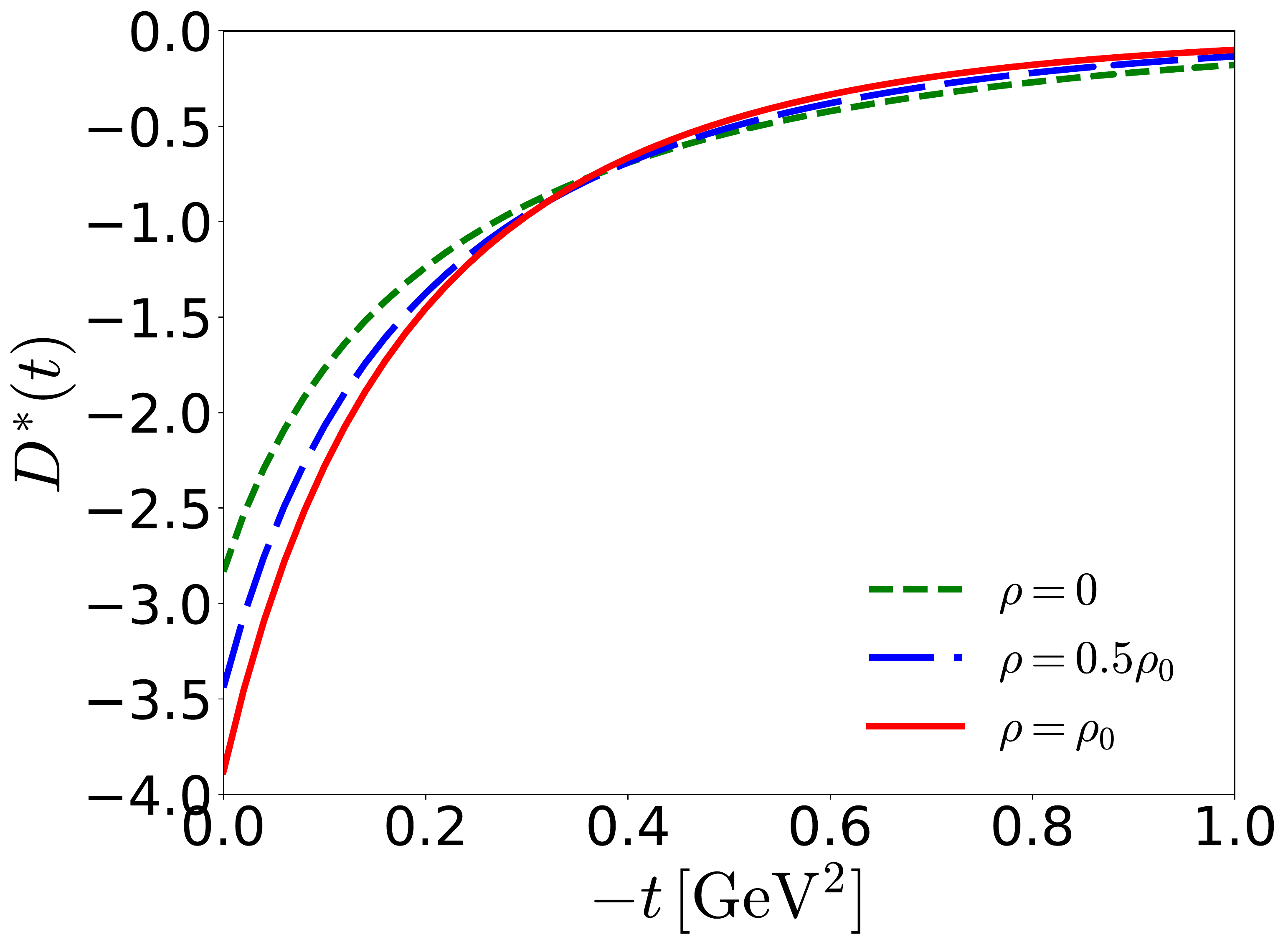}
\end{center}
\caption{The GFFs $A^{*}(t), J^{*}(t)$, and $D^{*}(t)$ of the nucleon
  as functions of the momentum transfer $t$. The short-dashed 
  curves draw those in free space, whereas the long-dashed and solid
  ones depict those respectively at $\rho=0.5\rho_{0}$ and at
  $\rho=\rho_{0}$, where $\rho_0=0.5\,m_\pi^3$ is the normal nuclear
  matter density.}  
\label{fig:1}
\end{figure*}
To examine how the mechanical properties of the nucleon undergo
modification in nuclear matter, we consider homogeneous nuclear matter
in which the nuclear density is kept constant. From now on,
all observables are given as functions of the normalized nuclear
matter density $\rho/\rho_{0}$. 

\subsection{Gravitational form factors of the nucleon}
The GFFs of the nucleon and the corresponding 3D densities in
nuclear matter were already studied in Ref.~\cite{Kim:2012ts}.
Since, however, the 3D distributions acquire ambiguous relativistic
corrections because of the relativistic nature of the nucleon, they do
not give a correct probabilistic meaning. Thus, we need to consider
the 2D distributions of the nucleon in the LF, which provide proper
probabilistic meaning. We can obtain the 2D
densities in the LF by using the Abel transformation. 

In Fig.~\ref{fig:1}, we show the results of the nucleon GFFs as
functions of $t$, changing the nuclear densities from $\rho=0$ (free
space) to $\rho=\rho_0$. As $\rho$ increases, The form factors 
fall off faster than those in free space as functions of $t$. 
Since $A^{*}(0)=1$ and $J^{*}(0)=1/2$ are
strictly constrained to be at $t = 0$, their magnitudes are
unchanged in nuclear matter. It hints that the corresponding radii of
the nucleon get larger in the medium, i.e., the size of the nucleon
swells in medium, as was discussed in Ref.~\cite{Kim:2012ts}.  
On the other hand, the absolute magnitude of the in-medium $D^{*}(t)$
is enhanced by the density effects but decreases more noticeably than
that in free space as $t$ increases. The $D$-term form factor contains
information on the stability of the nucleon and characterizes the
distribution of force fields inside the nucleon. Many theoretical
studies and experimental indications show that it should be
negative~\cite{Goeke:2007fp, Burkert:2018bqq, Kumericki:2015lhb,
  Ji:1997gm, Pasquini:2014vua,LHPC:2007blg} to secure the stability of
the nucleon. In the present work, $D(0)$ is around $-2.8$ in free
space, but it decreases (becomes larger in the absolute value) as
$\rho$ increases. $D^*(0)$ at the normal nuclear matter density yields 
around $-3.9$. This result indicates that the pressure and shear-force
densities are deformed inside the nucleon by the nuclear
environment. In this context, it is interesting to note that  
the analysis of selected nuclear isotopes with spin-parity quantum 
numbers $J^P = 0^+$ shows the increasing absolute value of
$D(0)$~\cite{Guzey:2005ba}. For heavier nuclei, the value of the
$D$-term predicted in the framework of the liquid drop model is
approximated as $D(0)\approx -0.246 A^{2.26}$, where $A$ is the mass
number of the nucleus\,\cite{Polyakov:2002yz}. For more discussion,  
we also refer to Ref.~\cite{Polyakov:2018zvc}.

\subsection{Energy and angular momentum distributions of 
  the nucleon} 

We now discuss the results for the energy distributions of the nucleon
in nuclear matter. In Table~\ref{tab:1}, various quantities related to
the GFFs of the nucleon are presented both in free space and in
nuclear matter at the different values of nuclear density $\rho$.
The 3D $\varepsilon^*(0)$ and 2D $\mathcal{E}^*(0)$ at the center of
the nucleon decrease if $\rho$ increases. Moreover, as listed in 
Table~\ref{tab:1}, the central value of the 3D distribution lessens
faster than that of the 2D case. For example, the central value of the
3D distribution decreases by about 50~\%, whereas it decreases by
about 36~\% in the 2D case.  
\begin{table}[h]
\setlength{\tabcolsep}{5pt}
\renewcommand{\arraystretch}{1.5}
\caption{Various quantities related to the GFFs of the
  nucleon in both the BF and IMF: the energy distributions at the
  center $(\varepsilon^{*}(0), \, \mathcal{E}^{*}(0))$, the pressure
  distributions at the center $(p^{*}(0), \, \mathcal{P}^{*}(0))$, nodal
  points of the pressures $(r^{*}_{0},\,(x^{*}_{\perp})_{0}))$ and the mean
  square radii of the mass, angular momentum and mechanical
  ($\langle r^{2} \rangle^{*}, \, \langle x^{2}_{\perp} \rangle^{*}$)
  at different values of nuclear density $\rho$.}  
\begin{center}
\begin{tabular}{c c c c  } 
\hline
\hline{}
$\rho/\rho_0$ & 0  & 0.5
  & 1  \\  
\hline 
$\varepsilon^{*}(0)$ ($\mathrm{GeV/fm}^{3}$)  & 1.25 &  0.84 & 0.62 \\ 
 $p^{*}(0)$
 ($\mathrm{GeV/fm}^{3}$) & 0.263 & $0.178$ & 0.133  \\ 
$r^{*}_{0}$ ($\mathrm{fm}$)  & 0.72 & 0.82 & 0.91   \\
$\langle r^{2}
  \rangle^{*}_{\mathrm{\varepsilon}}$ ($\mathrm{fm}^{2}$)&0.68&0.83&0.95\\
   $\langle r^{2}  \rangle^{*}_{J}$
 ($\mathrm{fm}^{2}$)&1.09&1.23&1.35\\
 $\langle r^{2} \rangle^{*}_{\mathrm{mech}}$ 
($\mathrm{fm}^{2}$) &0.75&0.90&1.01\\
 \hline 
$\mathcal{E}^{*}(0)$ ($\mathrm{GeV/fm}^{2}$) &1.25 & 0.96 & 0.80  \\
$\mathcal{P}^{*}(0)$
   ($\mathrm{GeV/fm}^{2}$) &0.060 & 0.047 & 0.039   \\
$(x_{\perp})^{*}_{0}$ ($\mathrm{fm}$)  &0.59 & 0.67 & 0.73 \\
$\langle x_{\perp}^{2}
 \rangle^{*}_{\mathcal{E}}$($\mathrm{fm}^{2}$)&
 0.30&0.38&0.44\\
 $\langle
 x_{\perp}^{2} \rangle^{*}_{J}$ ($\mathrm{fm}^{2}$)&0.87&0.99&1.08\\
 $\langle x_{\perp}^{2} \rangle^{*}_{\mathrm{mech}}$
    ($\mathrm{fm}^{2}$)&0.50&0.60&0.68\\
\hline
\hline
\end{tabular}
\end{center}
\label{tab:1}
\end{table}

\begin{figure*}[htp]
\begin{center}
\includegraphics[scale=0.26]{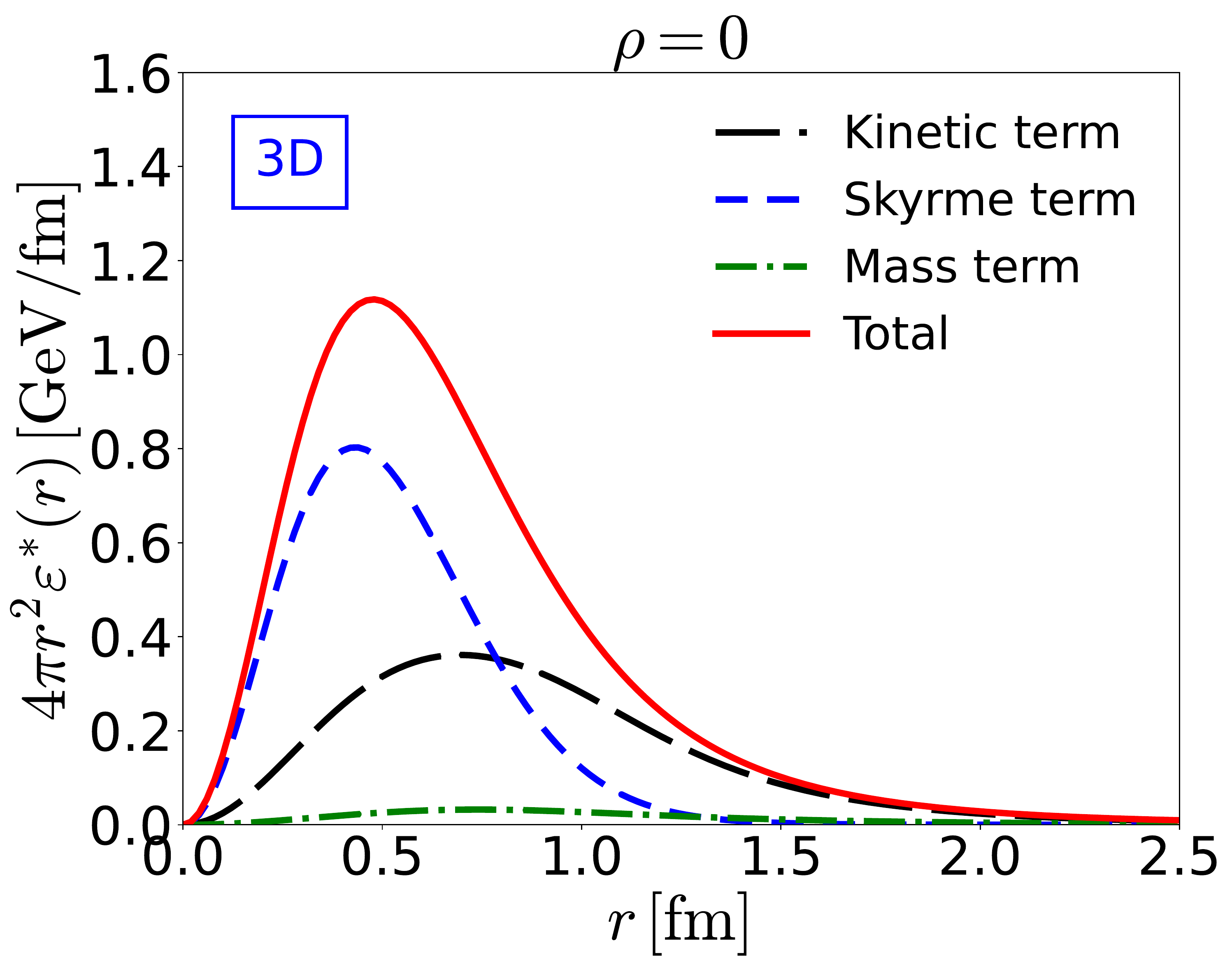}
\hspace{0.75cm}
\includegraphics[scale=0.26]{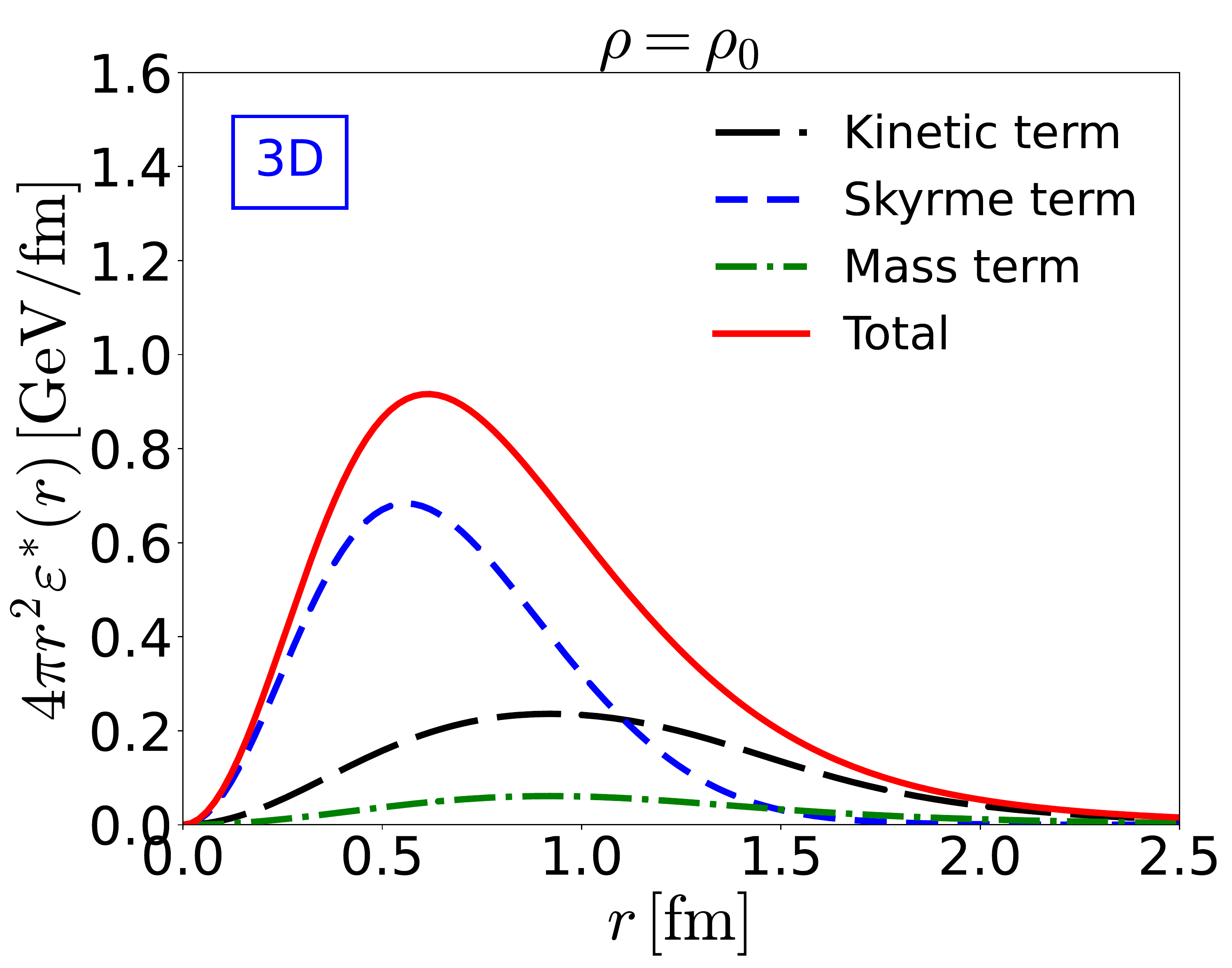}
\includegraphics[scale=0.26]{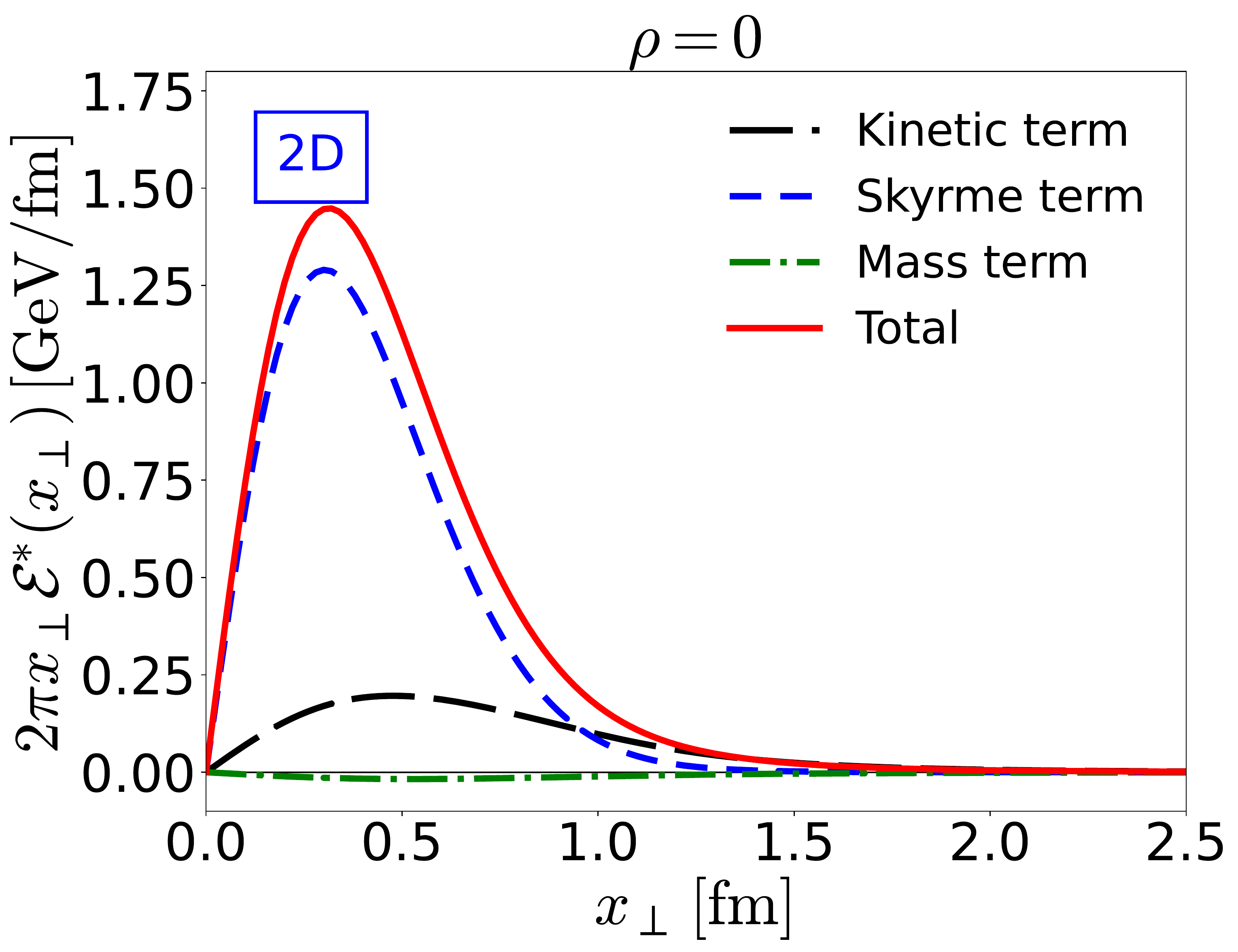}
\hspace{0.75cm}
\includegraphics[scale=0.26]{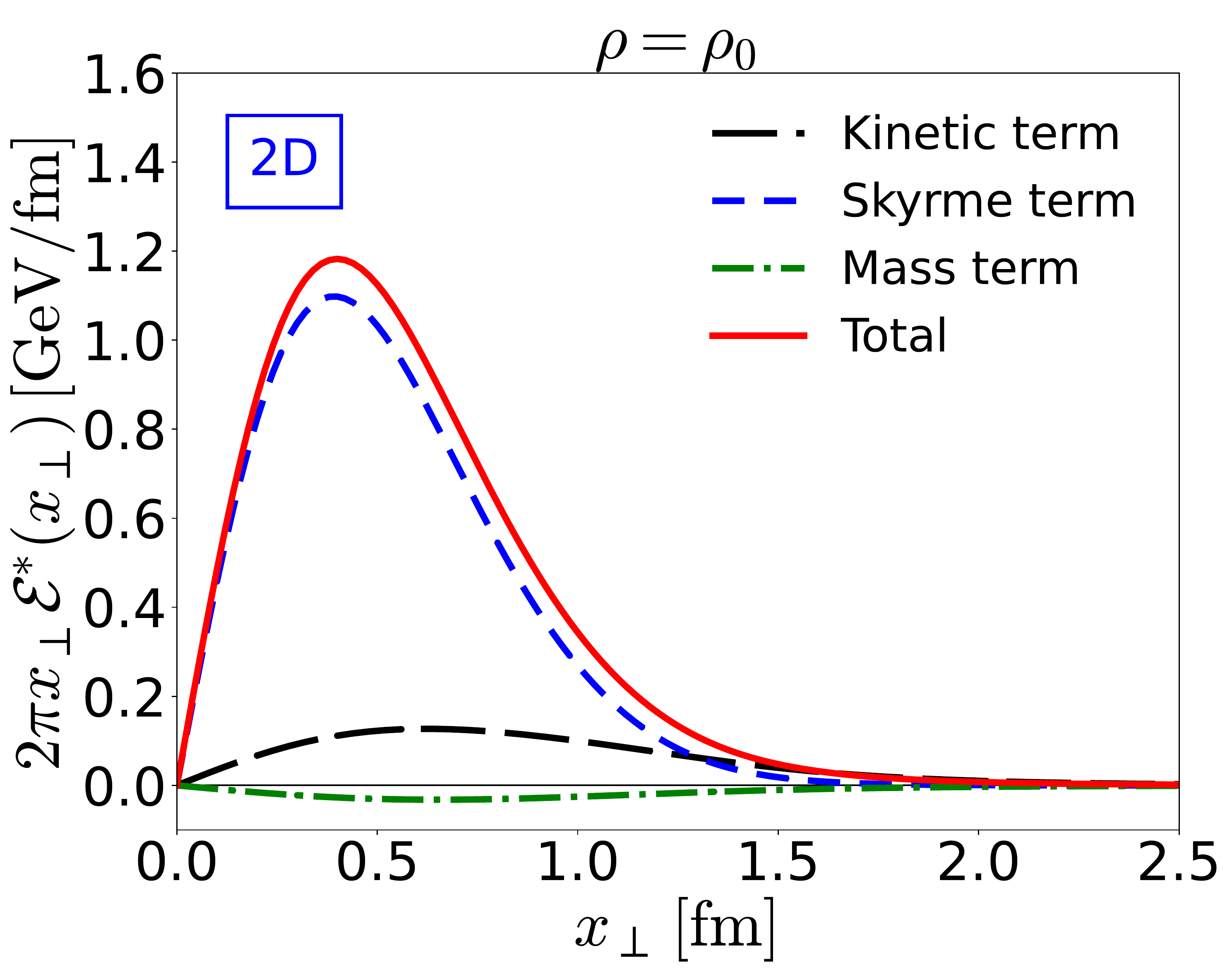}
\end{center}
\caption{The 3D and 2D mass distributions multiplied respectively by
  $4\pi r^{2}$ and $2\pi x_{\perp}$. The upper-left (-right) panel
  depicts 3D mass distributions with $4\pi r^{2}$ at $\rho=0$
  ($\rho=\rho_{0}$). The long-dashed, short-dashed, dot-dashed, and
  solid curves illustrate the contributions of the
  kinetic term with two derivatives, the Skyrme term with four
  derivatives, the mass term, and the total result, respectively. The
  lower-left (-right) panel draws the 2D mass distributions with $2\pi
  x_{\perp}$ at $\rho=0$   ($\rho=\rho_{0}$). Notations are the same as  
  in the 3D case.} 
\label{fig:2}
\end{figure*}
In Fig.~\ref{fig:2}, we draw the 3D and 2D mass distributions
multiplied respectively by $4\pi r^{2}$ and $2\pi x_{\perp}$. We
find that the 3D one exhibits a broader shape than the 2D one by
comparing the solid red curve in the upper panel with the lower panel. 
The 2D and 3D mass distributions become broader as the density of
nuclear matter increases, as seen from the solid red   
curve in the left panel compared to the right panel.   
These features appear clearly in the mean square radii of
the 2D and 3D mass distributions:   
\begin{align}
\frac{\langle x^{2}_{\perp} \rangle^{*}_{\mathcal{E}}}{\langle r^{2}
  \rangle^{*}_{\mathcal{\varepsilon}}} \approx \{0.441, 0.457 , 0.463
  \}  ,
\label{eq:r2}
\end{align}
where we list the ratios of mean square radii of the 2D and 3D energy
distributions given at three different values of the nuclear matter
density $\rho=\{0,0.5\rho_0,\rho_0\}$, respectively. In
Eq.~\eqref{eq:radius}, the 2D mass radius is expressed as $\langle
x_\perp^2\rangle_{\mathcal{E}}^* = \frac23 \langle
r^2\rangle_{\varepsilon}^* +D^*(0)/m^{*2}$. The factor $2/3$ comes from
the geometrical difference. In addition, $D(0)$ should be negative to 
ensure the stability of the nucleon. Thus, it is natural for $\langle
x_\perp^2\rangle_{\mathcal{E}}$ to be smaller than $\langle
r^2\rangle_{\varepsilon}$. We list the numerical values of the 3D and
2D radii explicitly in Table~\ref{tab:1}. As a result, 
the 3D mass mean square radius decreases by about 40~\% at the
normal nuclear matter density, while in the 2D case, it is reduced by
about 47~\%, compared to those in free space, respectively. On the
other hand, the central value of the 2D mass distribution gets larger
than the 3D one as the nuclear density increases. 

It is interesting to analyze how the different terms of the effective
chiral Lagrangian contribute to the mass distribution. As shown in
Fig.~\ref{fig:2}, all contributions become broader as the nuclear
density increases. It indicates that both the internal core of the
nucleon and outer shell are affected in the presence of the nuclear
environment. The main contribution to the mass comes from 
the Skyrme term, whereas that of the mass term turns out to be the
smallest. It is of particular interest to examine how each term 
contributes after the Abel transformation. While the contributions
from the kinetic and Skyrme terms preserve their sign, 
the mass distribution from the mass term becomes negative in both the
free space and nuclear matter after the Abel transformation.
This  behavior can be understood from Eq.~\eqref{eq:Abel_PS}. As will
be discussed later, mass term contributes to the pressure distribution
negatively, i.e., it gives $3p(r)/2<0$, of which the magnitude is
larger than $\varepsilon(r)$.  

\begin{figure*}[htp]
\begin{center}
\includegraphics[scale=0.265]{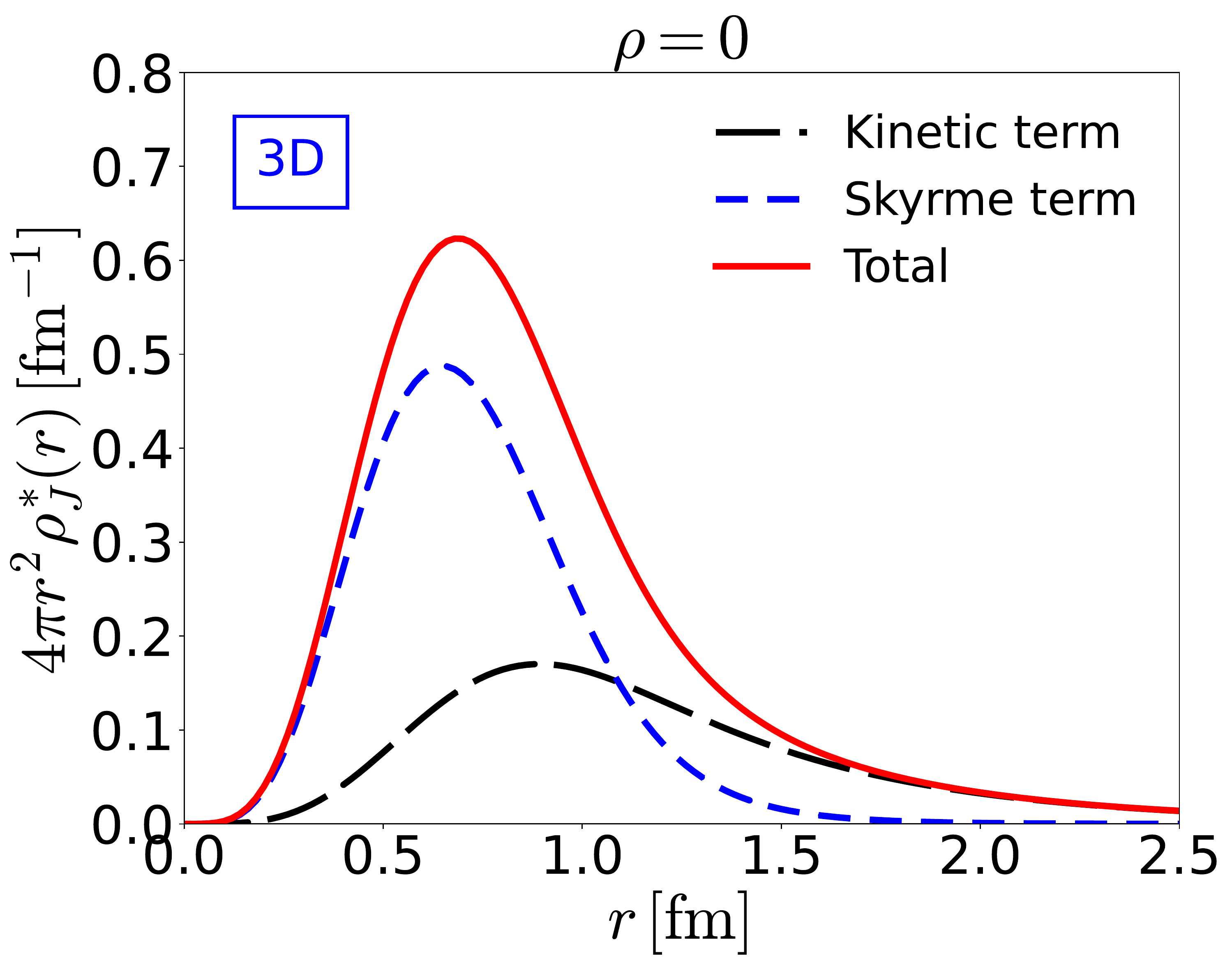}
\hspace{0.75cm}
\includegraphics[scale=0.265]{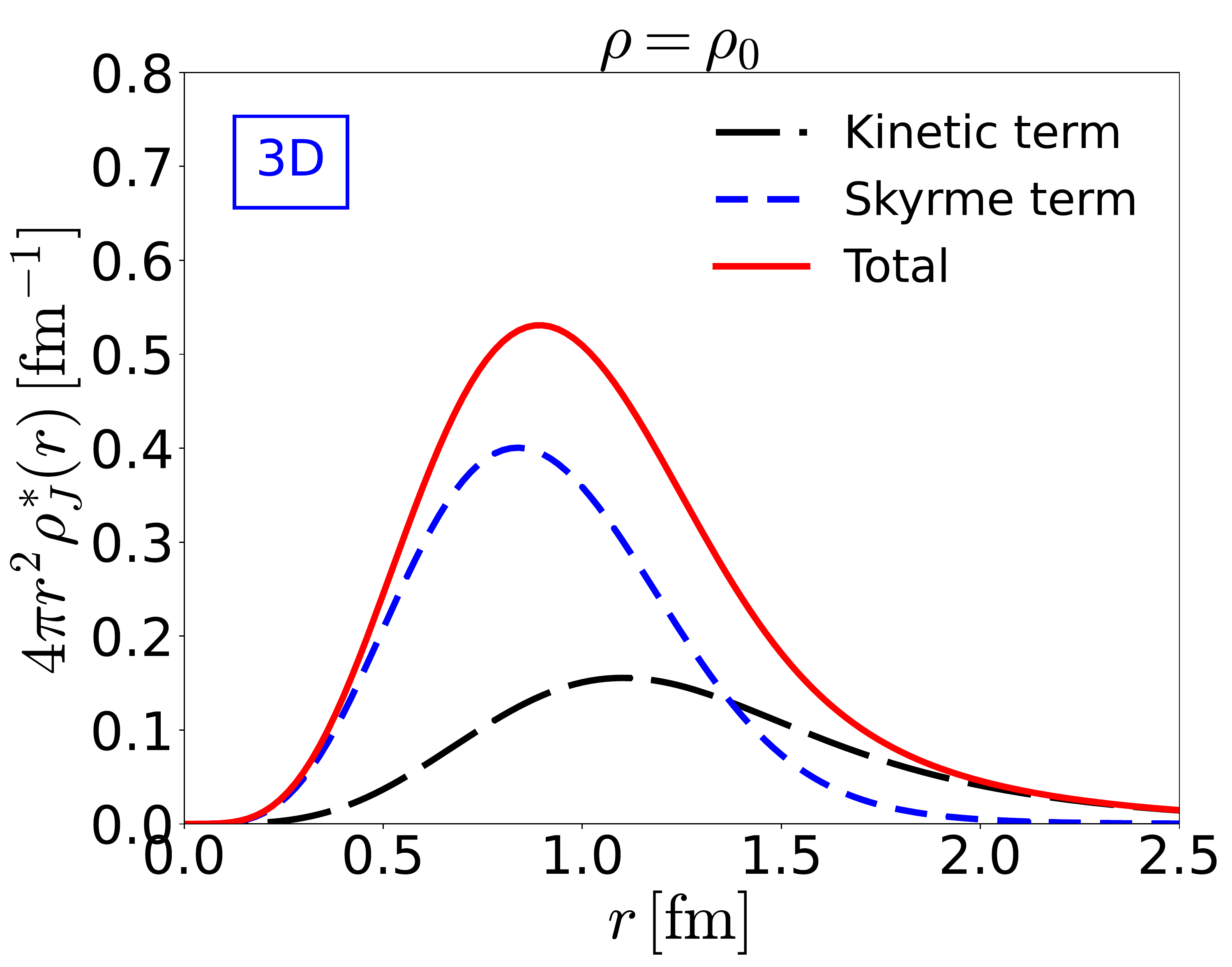}
\includegraphics[scale=0.265]{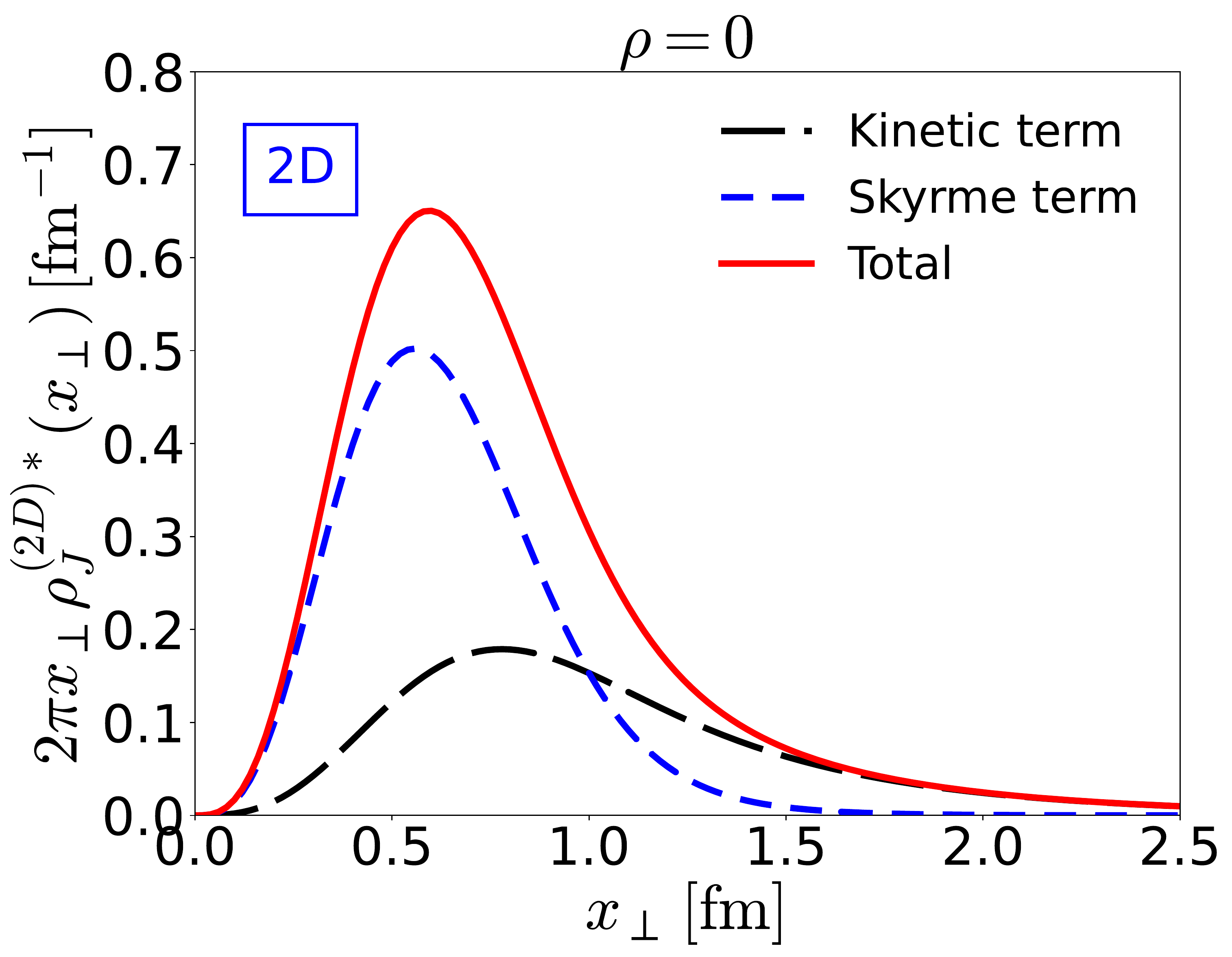}
\hspace{0.75cm}
\includegraphics[scale=0.265]{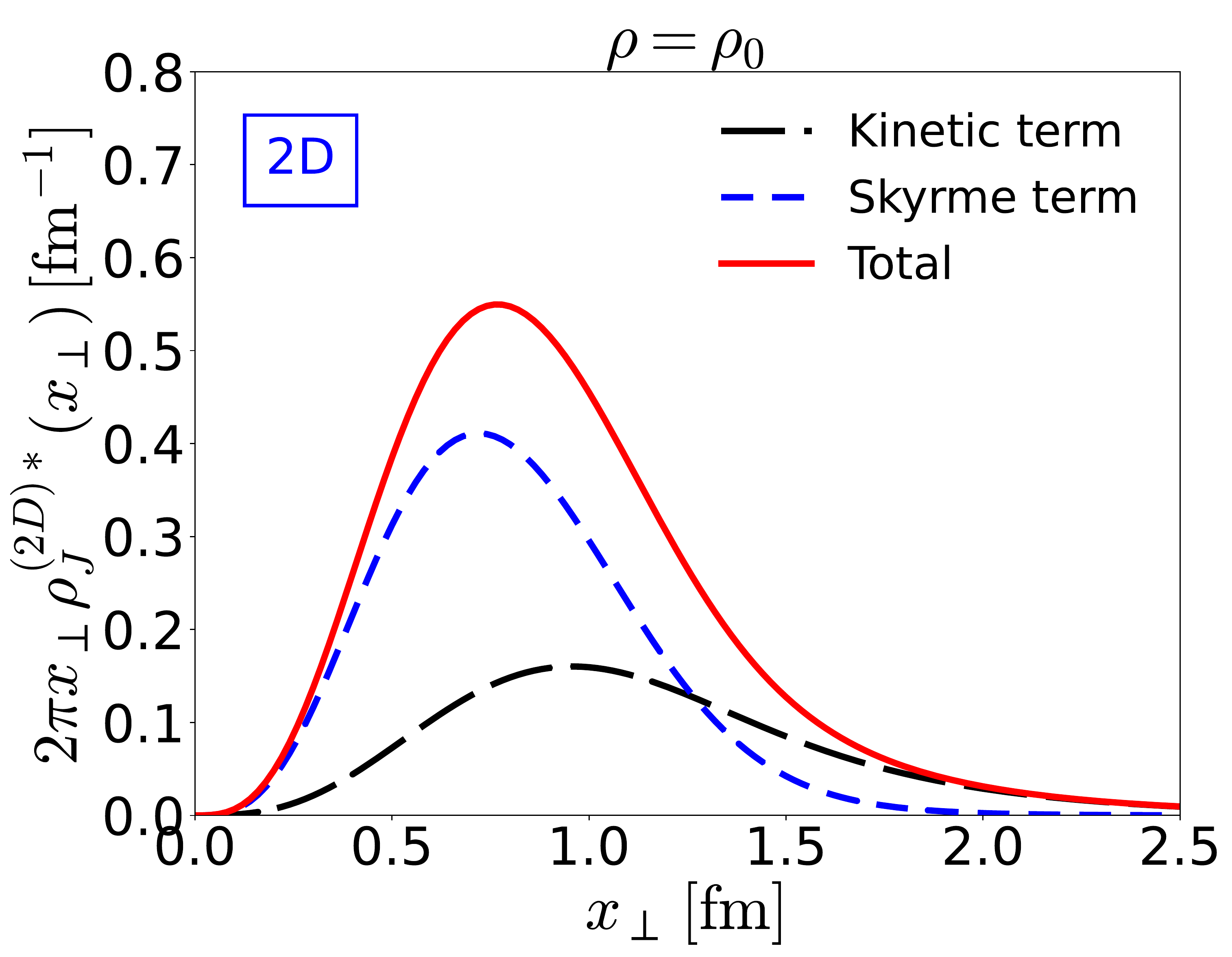}
\end{center}
\caption{The 3D angular momentum distributions of the nucleon. 
The upper-left (-right) panel depicts the 3D angular momentum
  distributions multiplied by $4\pi r^{2}$ at $\rho=0$ 
  ($\rho=\rho_{0}$). The lower-left (-right) panel draws the 2D angular
  momentum distributions with $2\pi x_{\perp}$ at $\rho=0$
  ($\rho=\rho_{0}$). Notations are the same as in Fig.~\ref{fig:2}.}  
\label{fig:3}
\end{figure*}
In Fig.~\ref{fig:3}, we depict the 3D and 2D distributions for the
angular momentum, multiplied by $4\pi r^{2}$ and $2\pi
x_{\perp}$, respectively. Note that integrating over the 2D and 3D
angular-momentum distributions should yield the spin of the nucleon
\begin{align}
\int d^{2}x_{\perp} \rho^{(2D)*}_{J}(x_{\perp}) = \int d^{3}r
  \rho^{*}_{J}(r) = J^{*}(0) = \frac{1}{2}, 
\end{align}
which provides a self-consistency check. As in the case of the mass
distributions, both the 3D and 2D distributions for the angular
momentum become broader as the nuclear density increases.
It implies that the mean square radii of the angular momentum 
distribution also increase in nuclear matter (see also
Table~\ref{tab:1}). The 3D distribution is broader than
the 2D one. The factor $4/5$ in Eq.~\eqref{eq:radius} arises again
from the geometrical difference between the 3D and 2D angular-momentum 
distributions. 
The sign of each contribution is not changed after the Abel
transformation. As shown in Eq.~\eqref{eq:Abel_PS},
$\rho_J^{(2D)}(x_\perp)$ is directly related to $\rho_J(r)$. 
As listed in Table~\ref{tab:2}, each contribution to the spin of the
nucleon in nuclear matter is also changed. It is of great interest to
see that the contribution of the kinetic term is reduced in nuclear
matter whereas that of the Skyrme term is enhanced as the nuclear
density increases. Nevertheless, the spin of the nucleon is always
kept to be 1/2 as it should be. 

\begin{table}[hb]
\setlength{\tabcolsep}{5pt}
\renewcommand{\arraystretch}{1.5}
\caption{
Contributions of the different terms to the angular
momentum $J$, the stability condition (Von Laue condition), and the
values of the $D$-term extracted from the pressure distributions for
$\rho=0$ ($\rho=\rho_0$).   
} 
\begin{tabular}{c l | c c c c c } 
\hline
\hline
\multicolumn{2}{c |}{Terms} &   $J$ & Von Laue & $D$-term \\ 
\hline
\multirow{3}{*}{3D} & $\mathcal{L}^*_2$ & 0.188 (0.179) & $-$9.55 ($-$7.74) & $-$2.78 ($-$3.26) \\
 & $\mathcal{L}^*_4$ &  0.312 (0.321) & 12.45 (13.65) & 1.16 (2.12)  \\
 & $\mathcal{L}^*_{\rm m}$  & $-$ & $-$2.90 ($-$5.91) &$-$1.22 ($-$2.74) \\
\hline 
\multirow{3}{*}{2D} &$\mathcal{L}^*_2$ & 0.188 (0.179) & $-$4.78 ($-$3.87) & $-$2.66 ($-$3.21)  \\ 
 &$\mathcal{L}^*_4$ & 0.312 (0.321) & 6.22 (6.83)  & 0.64 (1.15)\\ 
 &  $\mathcal{L}^*_{\rm m}$ & $-$& $-$1.44 ($-$2.96) & $-$0.82 ($-$1.82) \\
\hline 
 \multicolumn{2}{c|}{ Total}  & 0.5& 0 & $-$2.83 ($-$3.88)  \\ 
\hline 
\hline
\end{tabular}
\label{tab:2}
\end{table}
\subsection{Pressure and shear-force distributions of the nucleon}
\begin{figure*}[htp]
\begin{center}
\includegraphics[scale=0.265]{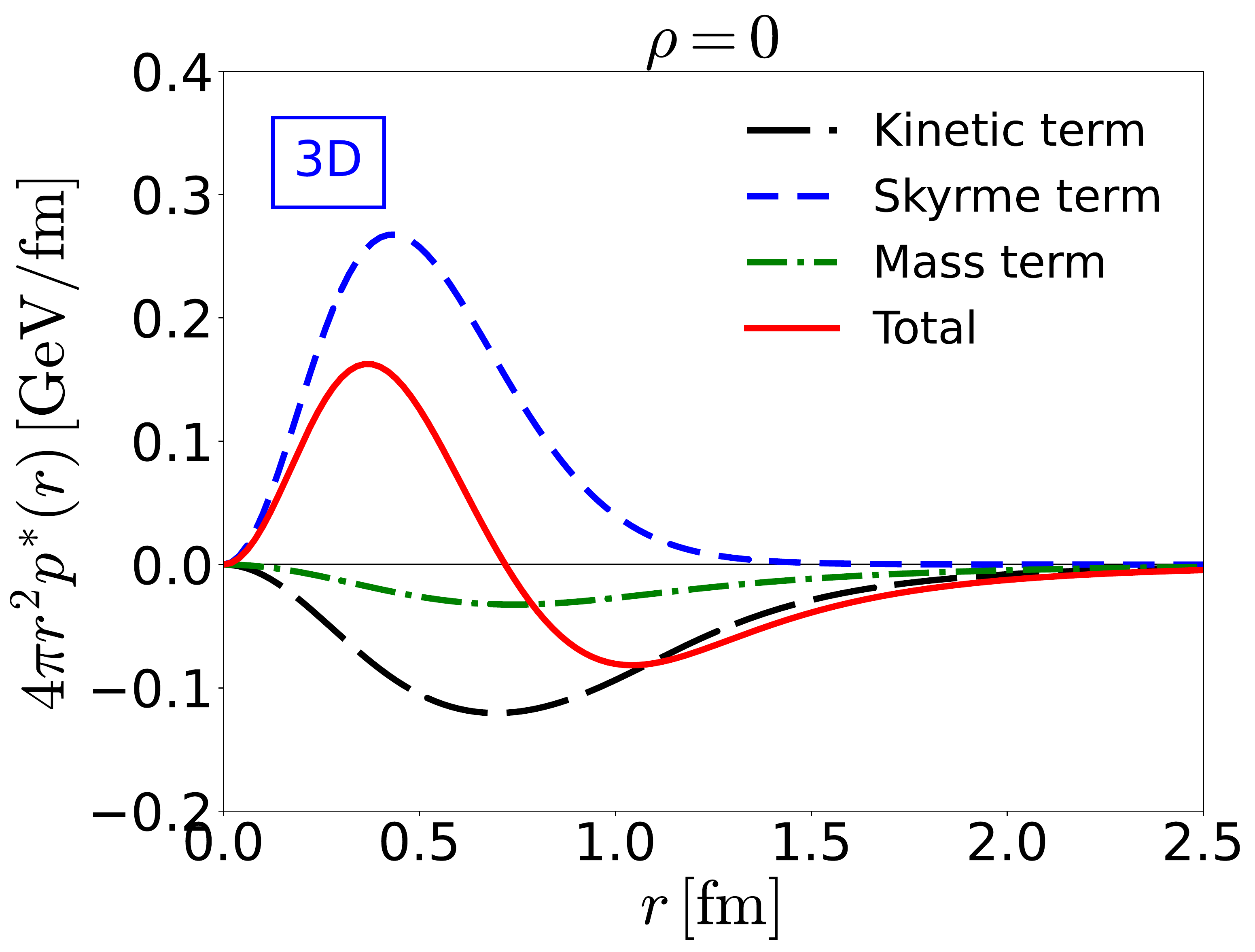}
\hspace{0.5cm}
\includegraphics[scale=0.265]{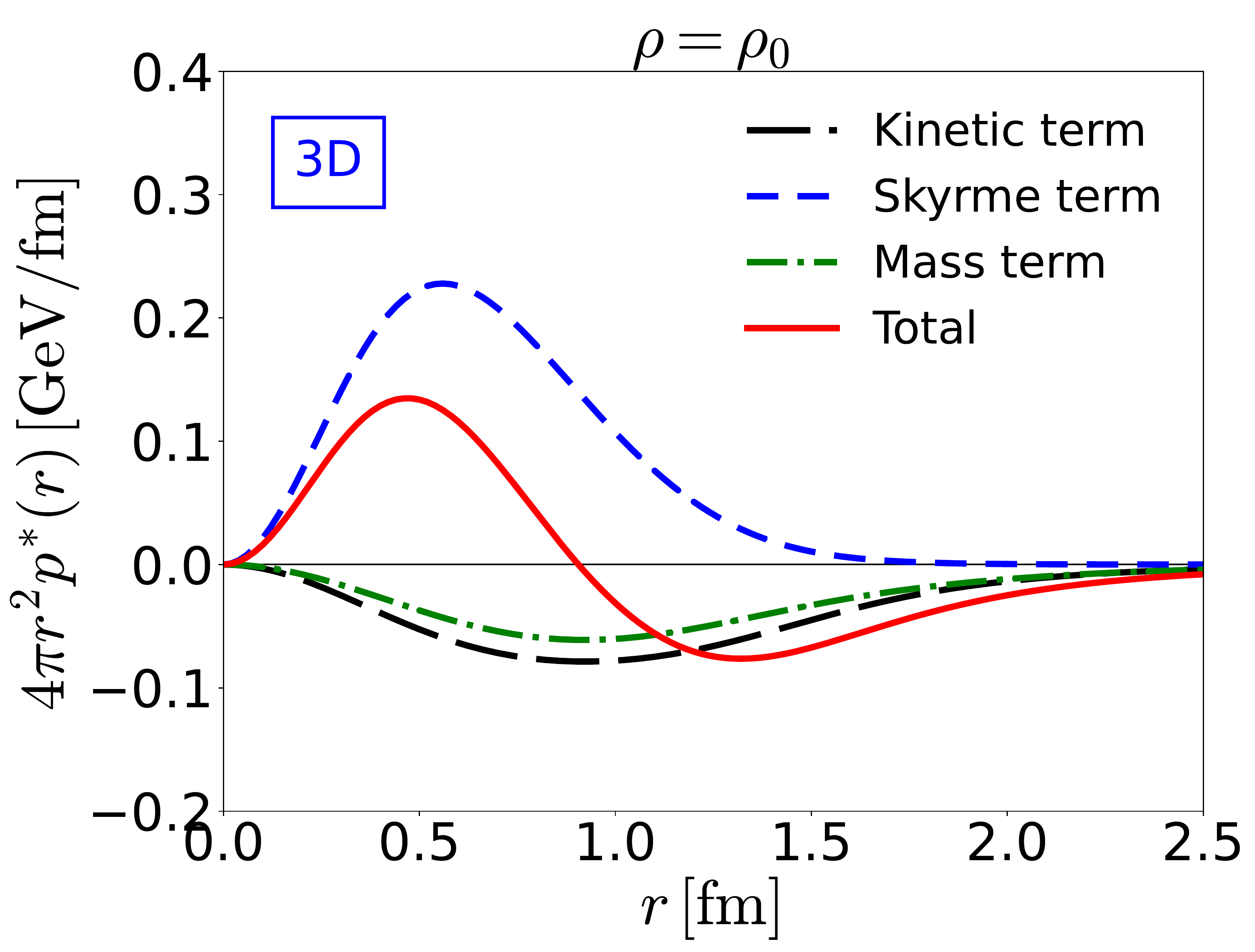}
\includegraphics[scale=0.265]{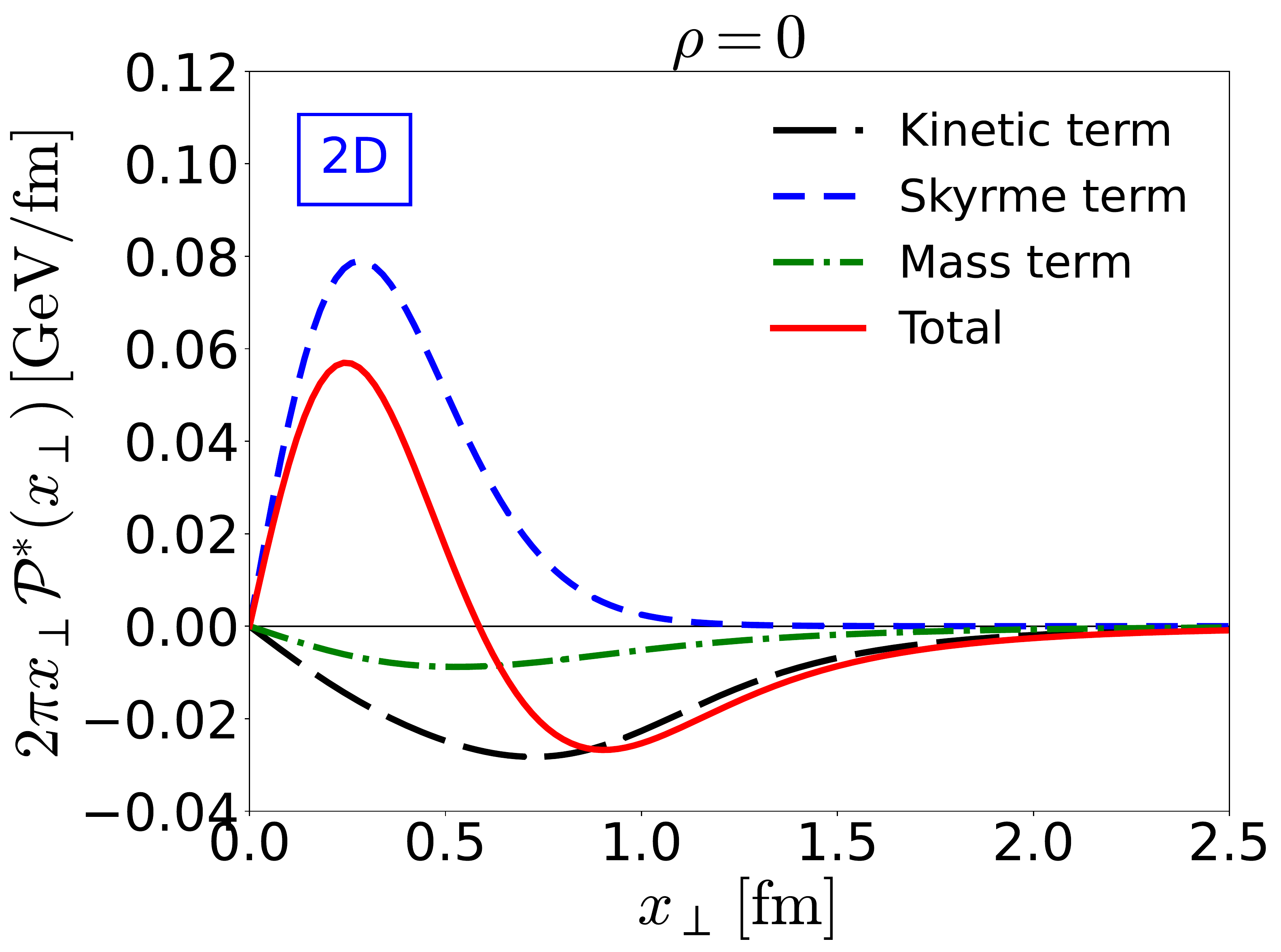}
\hspace{0.5cm}
\includegraphics[scale=0.265]{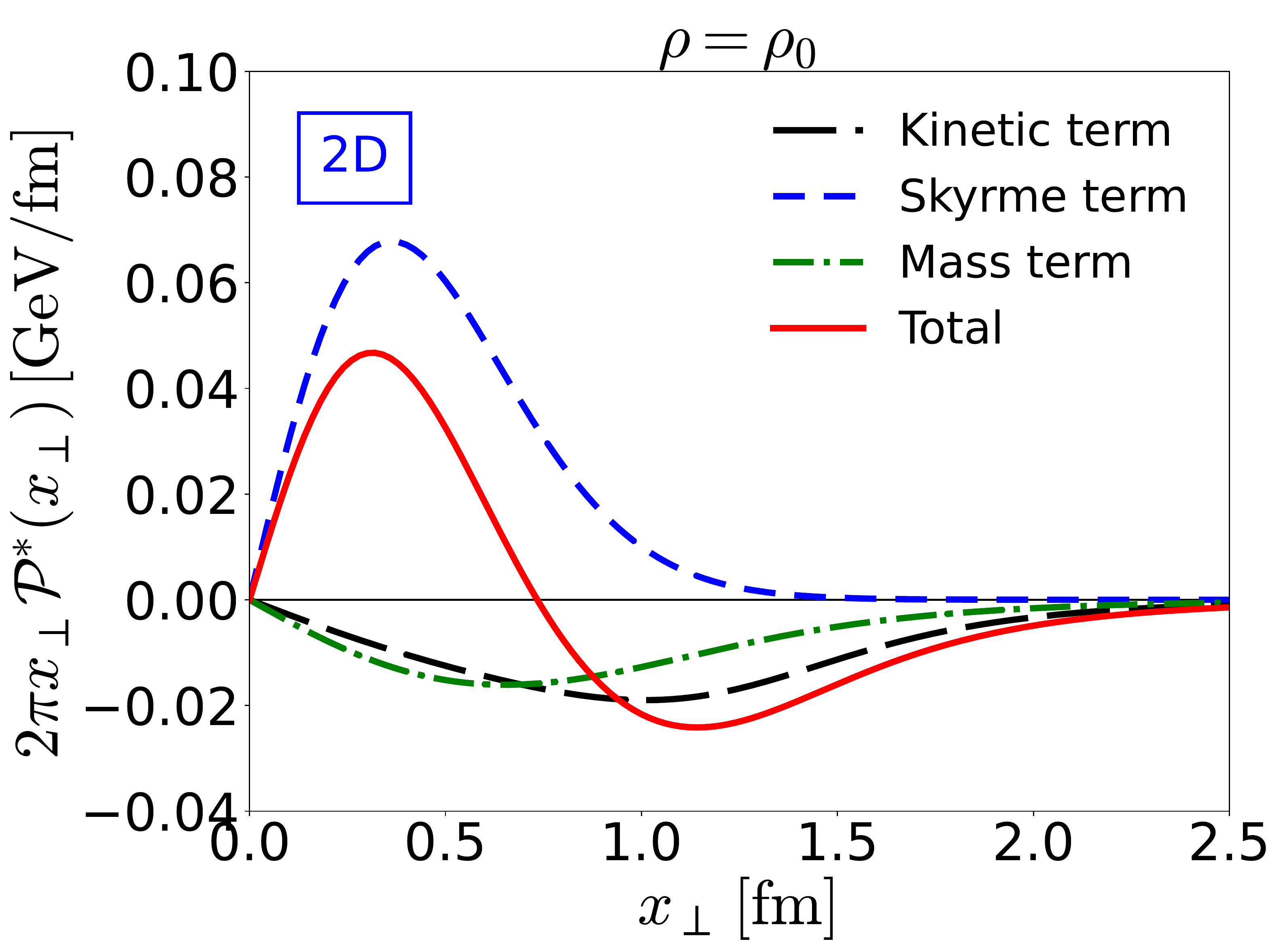}
\caption{3D and 2D pressure distributions of the nucleon.
The upper-left (right) panel depicts the 3D pressure
  distributions multiplied by $4\pi r^{2}$ at $\rho=0$ 
  ($\rho=\rho_{0}$). The lower-left (-right) panel draws the 2D
  pressure distributions with $2\pi x_{\perp}$ at $\rho=0$
  ($\rho=\rho_{0}$). Notations are the same as in Fig.~\ref{fig:2}.
}
\label{fig:4}
\end{center}
\end{figure*}

\begin{figure*}[htp]
\begin{center}
\includegraphics[scale=0.265]{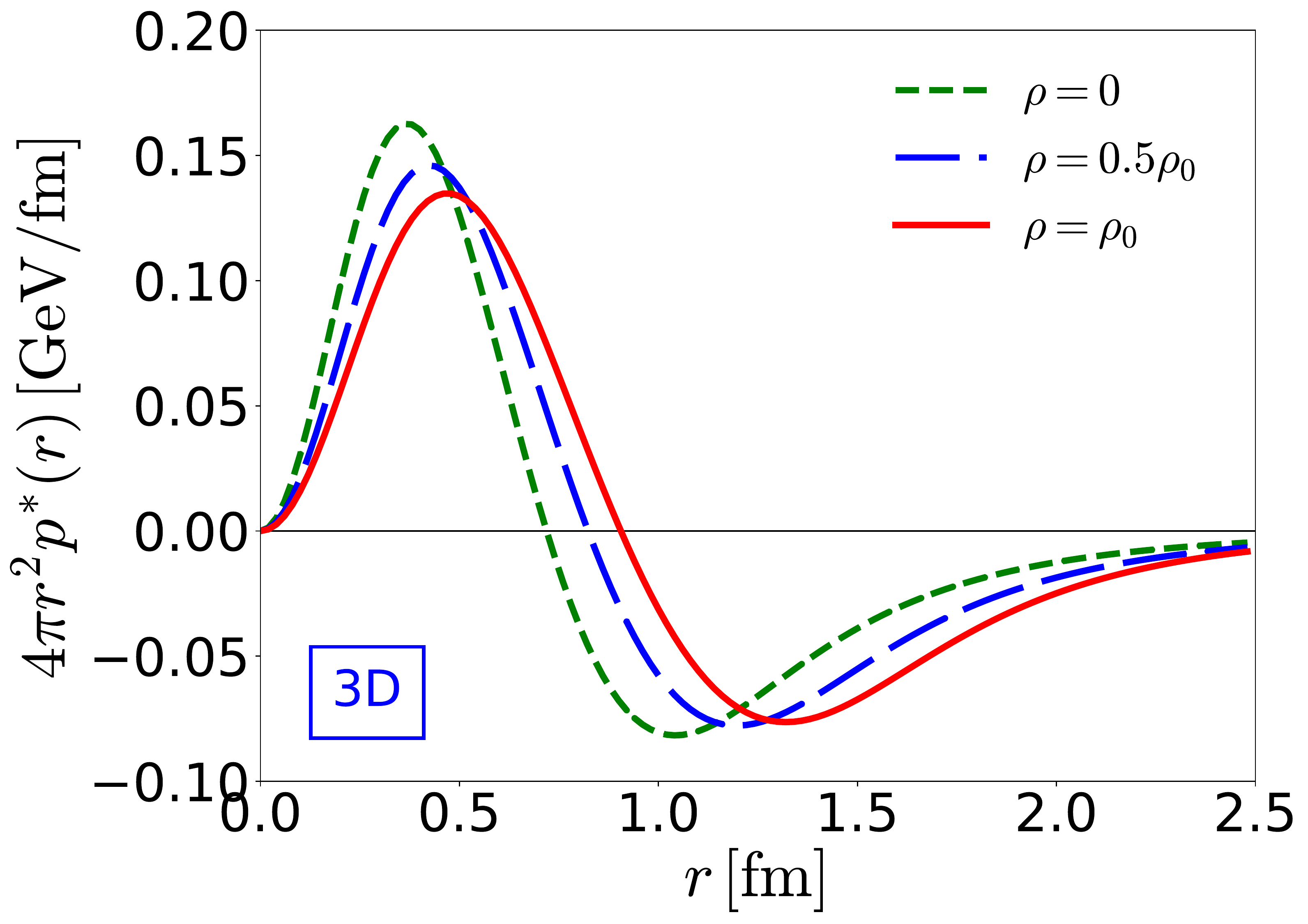}
\hspace{0.5cm}
\includegraphics[scale=0.265]{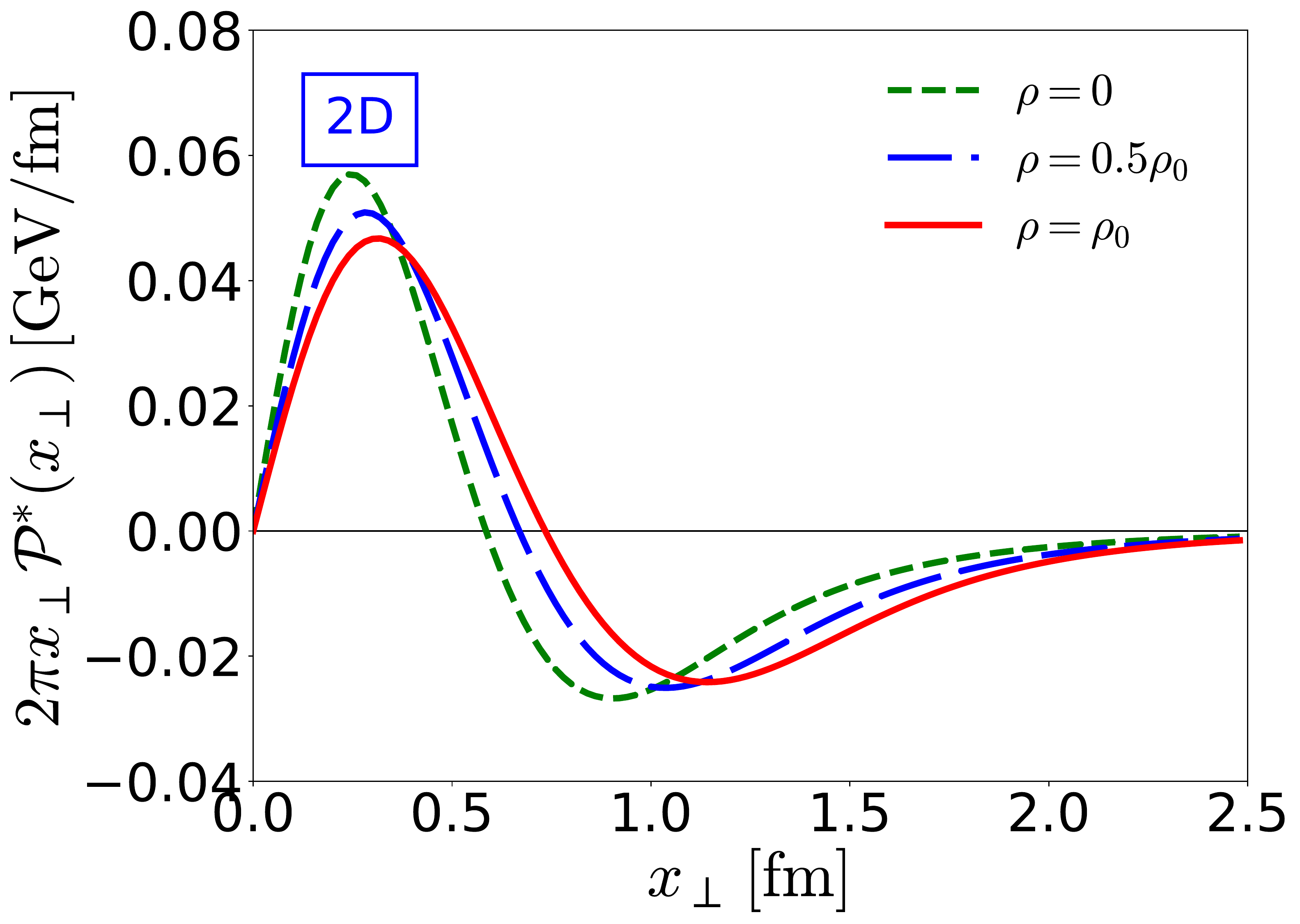}
\caption{3D and 2D pressure distributions of the nucleon 
as functions of the nuclear density $\rho$, multiplied by $4\pi r^2$
and $2\pi x_\perp$, respectively. The short-dashed,
long-dashed, and solid curves represent the pressure distributions
with $\rho=0,\,0.5\rho_0, \rho_0$, respectively. $\rho_0$ denotes the
normal nuclear matter density.  
}
\label{fig:5}
\end{center}
\end{figure*}
We are now in a position to discuss the pressure and shear-force
distributions, which are crucial for the stability
conditions of the nucleon. Figure~\ref{fig:4} shows each contribution
to the 3D and 2D pressure distributions multiplied by $4\pi r^2$ and
$2\pi x_\perp$ in the upper and lower panels, respectively. 
We draw the 3D and 2D weighted pressure densities in free space
(nuclear matter) in the upper-left (upper-right) and lower-left
(lower-left) panels, respectively. In fact, the results 
for the 3D distributions were already discussed in
Refs.~\cite{Cebulla:2007ei, Kim:2012ts}. Thus, we will concentrate in
this work on how the 2D pressure distributions undergo modification in
nuclear matter in comparison with the 3D ones. As shown in
Fig.~\ref{fig:4}, the Abel transformation does not change the general
feature of each contribution. It means that the conclusions drawn in
Refs.~\cite{Cebulla:2007ei, Kim:2012ts} are unchanged also in the 2D
pressure distributions. The kinetic and mass terms of the pion
provide the attractive force, whereas the Skyrme term yields the
repulsive force so that the stability of the nucleon is acquired. The
pressure distribution should have at least one nodal point such that
the global stability condition or the von Laue condition given in
Eq.~\eqref{eq:stabp} is satisfied. It is achieved by the perfect
balance between the kinetic and mass terms and the Skyrme term. 
In the case of the 2D pressure distribution, we have the same
conclusion. As listed in Table~\ref{tab:2}, the contributions of the
kinetic term to the 3D and 2D pressures decrease with the nuclear
density increased, whereas the absolute magnitudes of the Skyrme and
mass terms are enhanced in nuclear matter.   

In Fig.~\ref{fig:5}, we depict the 3D and 2D pressure distributions of
the nucleon as functions of the nuclear density $\rho$, multiplied by
$4\pi r^2$ and $2\pi x_\perp$, respectively. The results show clearly
that as $\rho$ increases, both the 3D and 2D pressure distributions
become broader. It implies that the nucleon swells in nuclear matter. 
If $\rho$ continuously increases to higher densities, 
the negative pressure distribution of the kinetic term gets further
shallowed, which ends up with the situation that at a certain nuclear
density the contribution of the kinetic term will no more 
compensate that of the Skyrme term. As a result, there is no
topological solution; the skyrmion disappears (for more
discussion, we refer to Ref.~\cite{Kim:2012ts}). It indicates that the
nucleon may dissolve in quark matter. 

The numerical value of $D(0)$ is obtained by integrating the 3D and 2D
pressure distributions with the weights $m r^2$ and $4m x_\perp^2$ as
shown in Eq.~\eqref{eq:dterm}. The weights amplify the pressure
distributions as $r$ and $x_\perp$ increase. Thus, the negative tails
of both $4\pi r^2 p^*(r)$ and $2\pi x_\perp \mathcal{P}^*(x_\perp)$
overwhelm the core parts so that the negative value of $D(0)$ is
ensured. It implies that the correct balance of the pressure
distributions is required to get the negative values of the
$D$-term. In Table~\ref{tab:2}, we list the results for the $D$-term
in the fourth column.  

\begin{figure*}[htp]
\begin{center}
\includegraphics[scale=0.265]{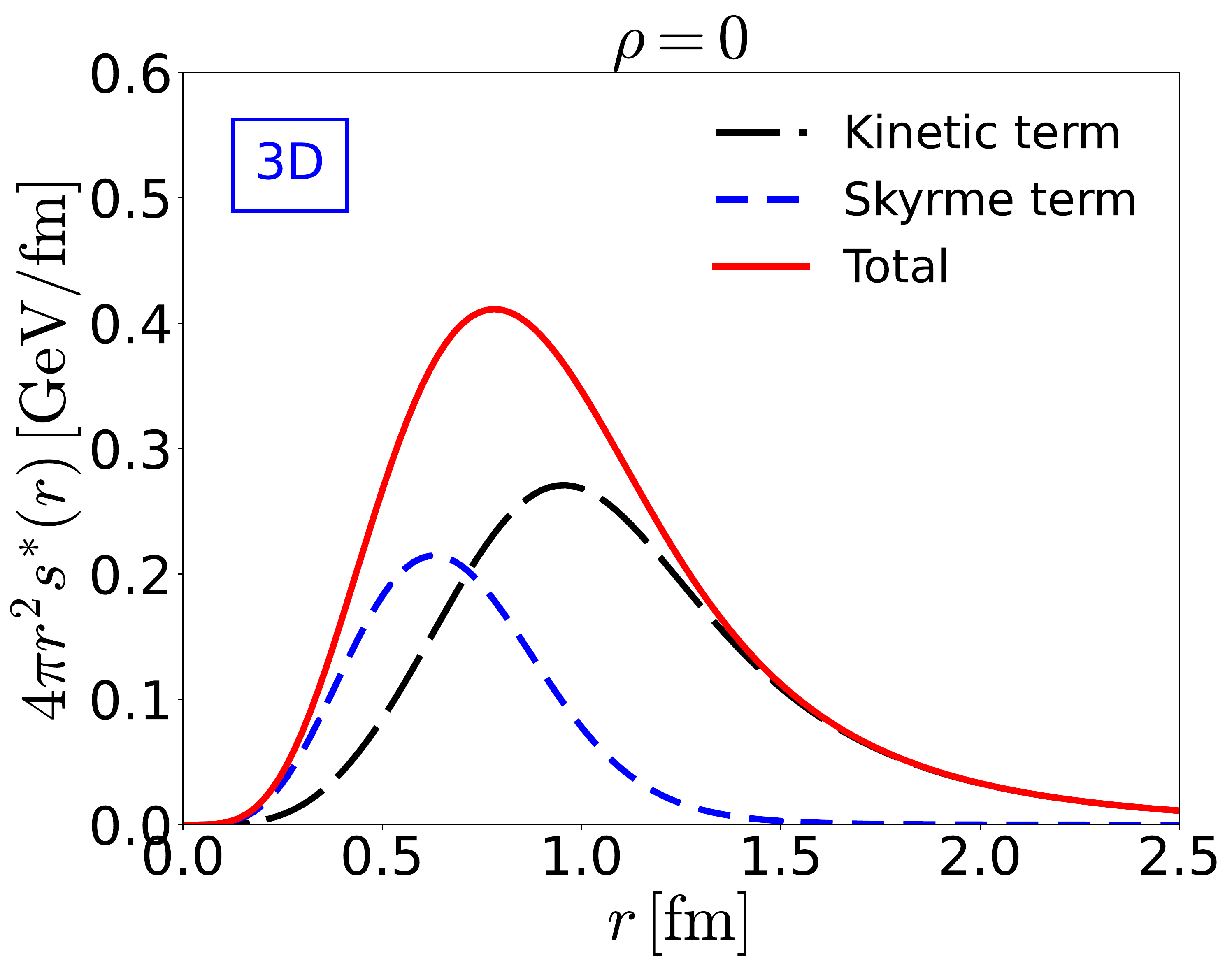}
\hspace{0.5cm}
\includegraphics[scale=0.265]{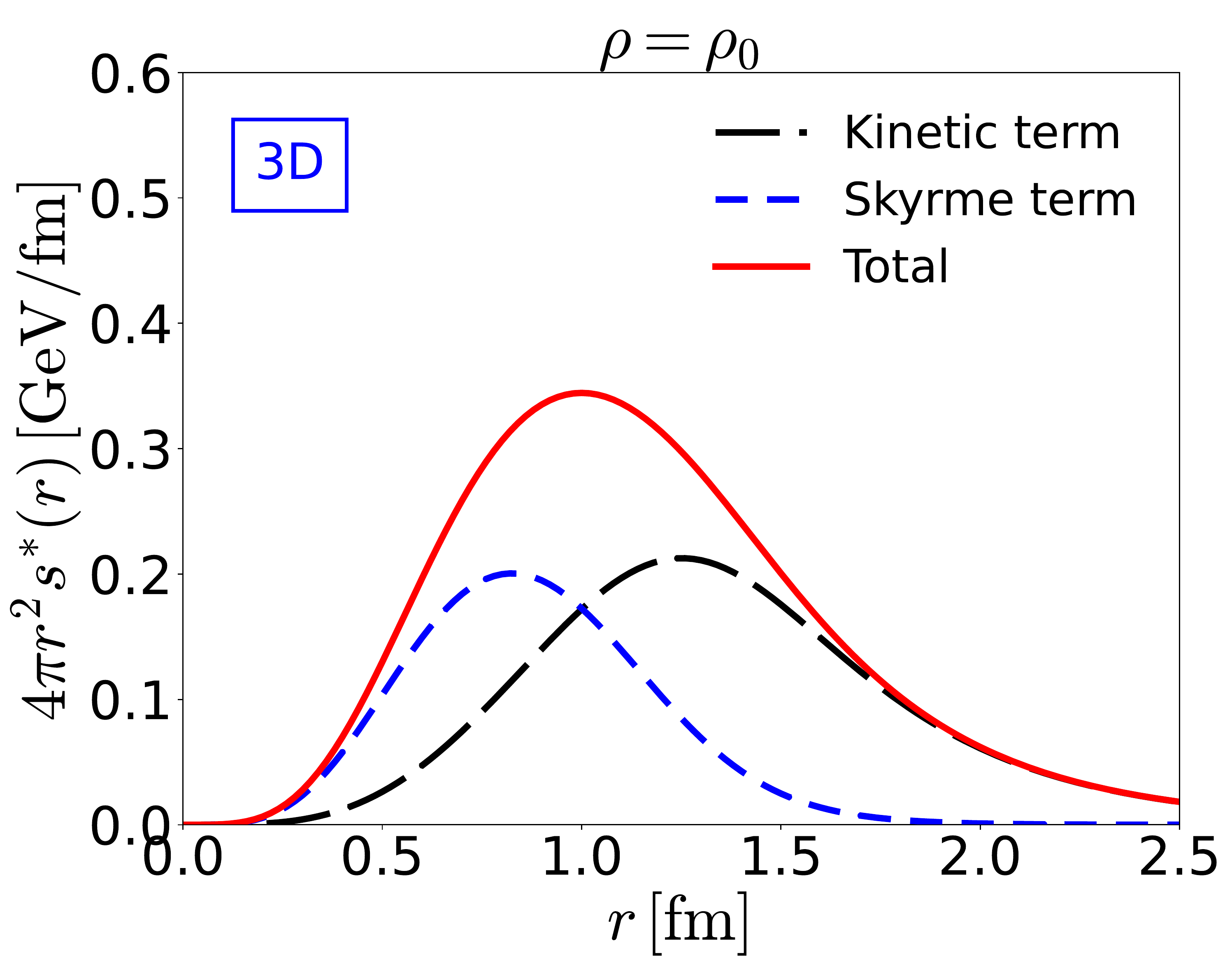}
\includegraphics[scale=0.265]{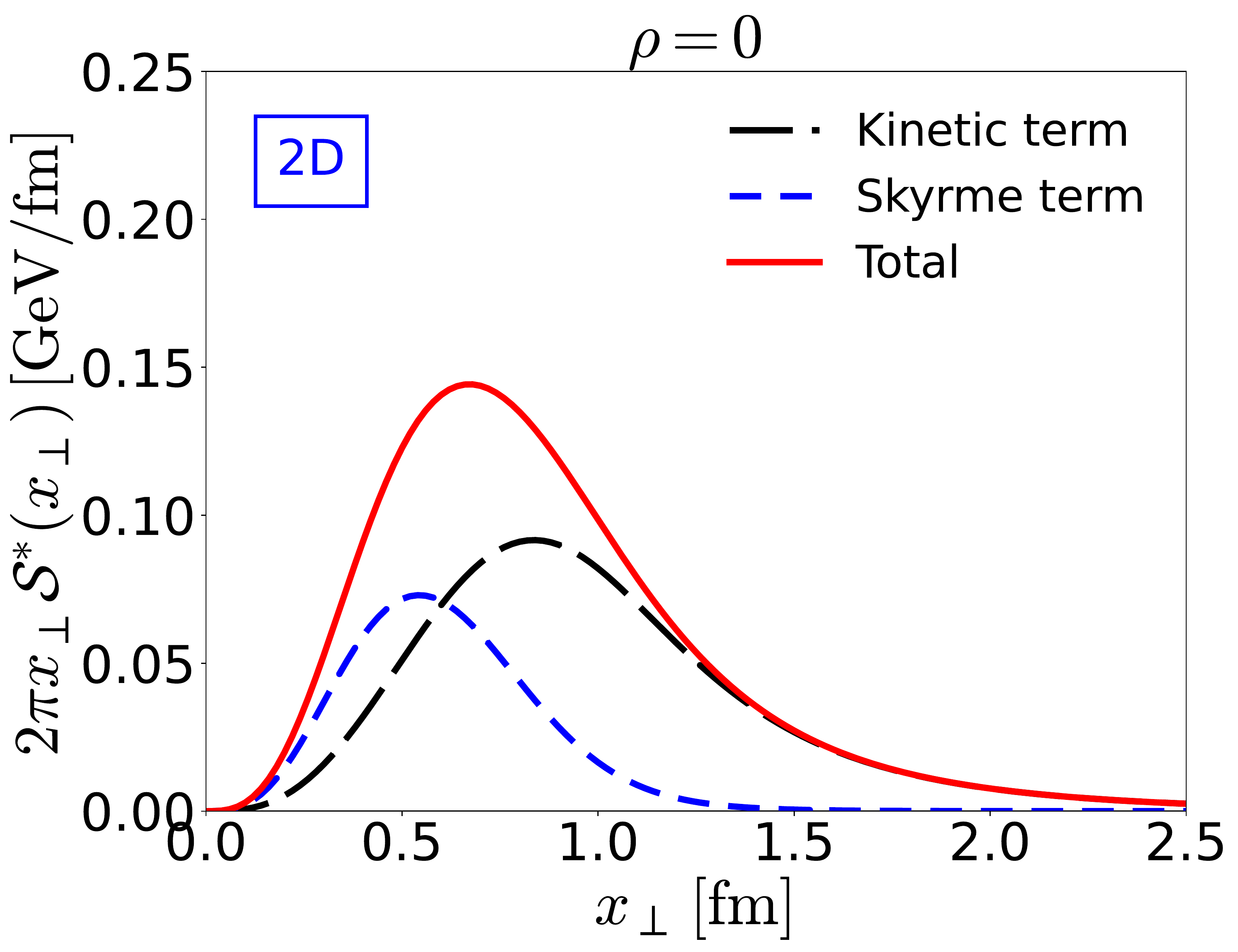}
\hspace{0.5cm}
\includegraphics[scale=0.265]{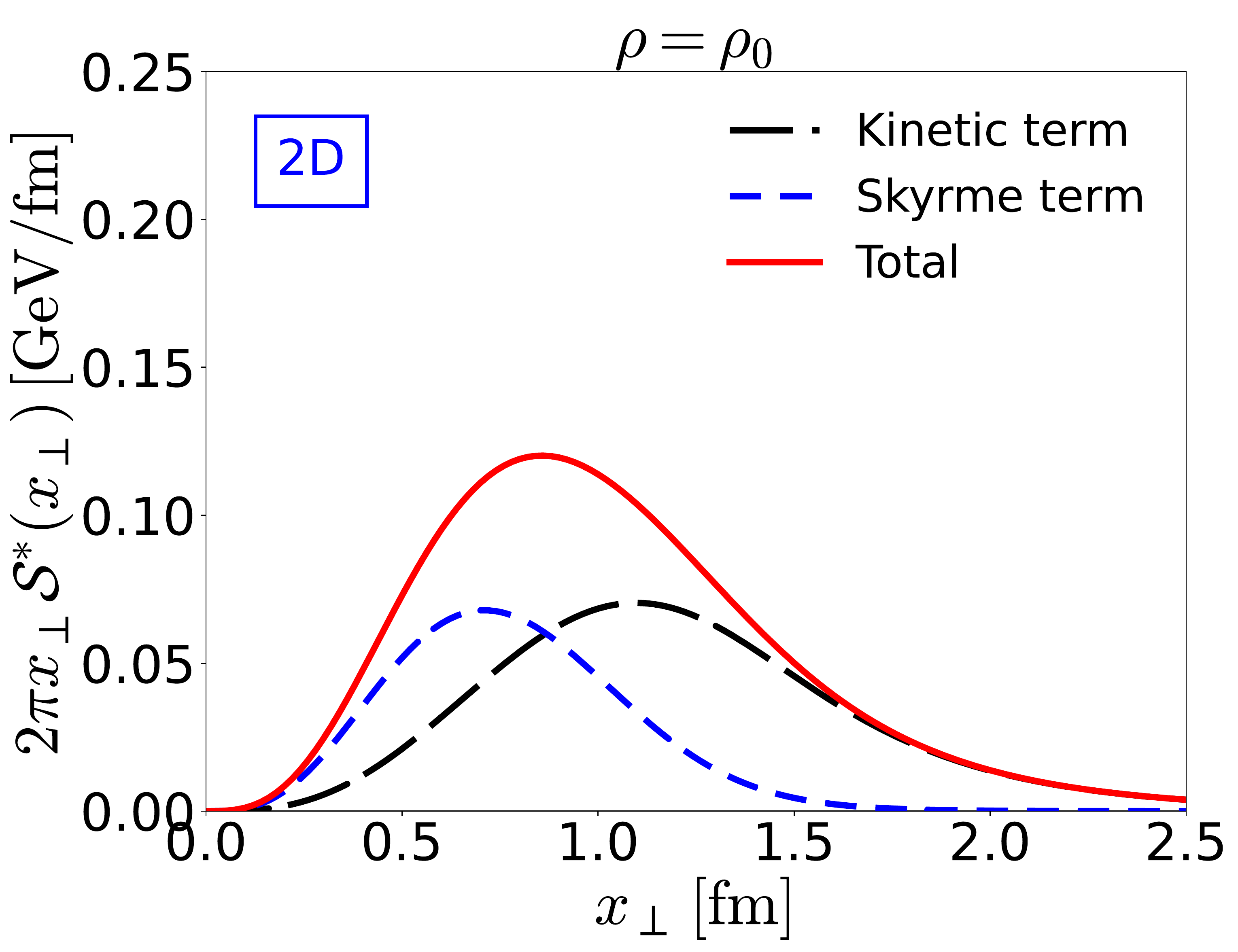}
\caption{
3D and 2D shear-force distributions of the nucleon.
The upper left (right) panel depicts the 3D shear-force
  distributions multiplied by $4\pi r^{2}$ at $\rho=0$ 
  ($\rho=\rho_{0}$). The lower left (right) panel draws the 2D
  shear-force distributions with $2\pi x_{\perp}$ at $\rho=0$
  ($\rho=\rho_{0}$). Notations are the same as in Fig.~\ref{fig:2}.
}
\label{fig:6}
\end{center}
\end{figure*}
Figure~\ref{fig:6} represents each contribution to the 3D and 2D
shear-force distributions of the nucleon multiplied by $4\pi r^2$ and
$2\pi x_\perp$ in the upper and lower panels, respectively. 
We want to mention that the mass term does not contribute to the
shear-force distribution as shown in Eq.~\eqref{eq:distributions}. 
The general feature of each contribution is again
well kept after the Abel transformation. In the case of the
shear-force distributions, the contribution of the kinetic term is
generally dominant over that of the Skyrme term. As shown in
Eq.~\eqref{eq:Abel_PS}, $S^*(x_\perp)$ is directly related to
$s^*(r)$, so that the shapes of the 3D shear-force distributions are 
maintained by the Abel transformation. Since both the 3D (2D)
shear-force distributions are positive over the whole ranges of $r$
($x_\perp$), the negative value of the $D$-term is also secured by
Eq.~\eqref{eq:dterm}. 

\begin{figure*}[htp]
\begin{center}
\includegraphics[scale=0.27]{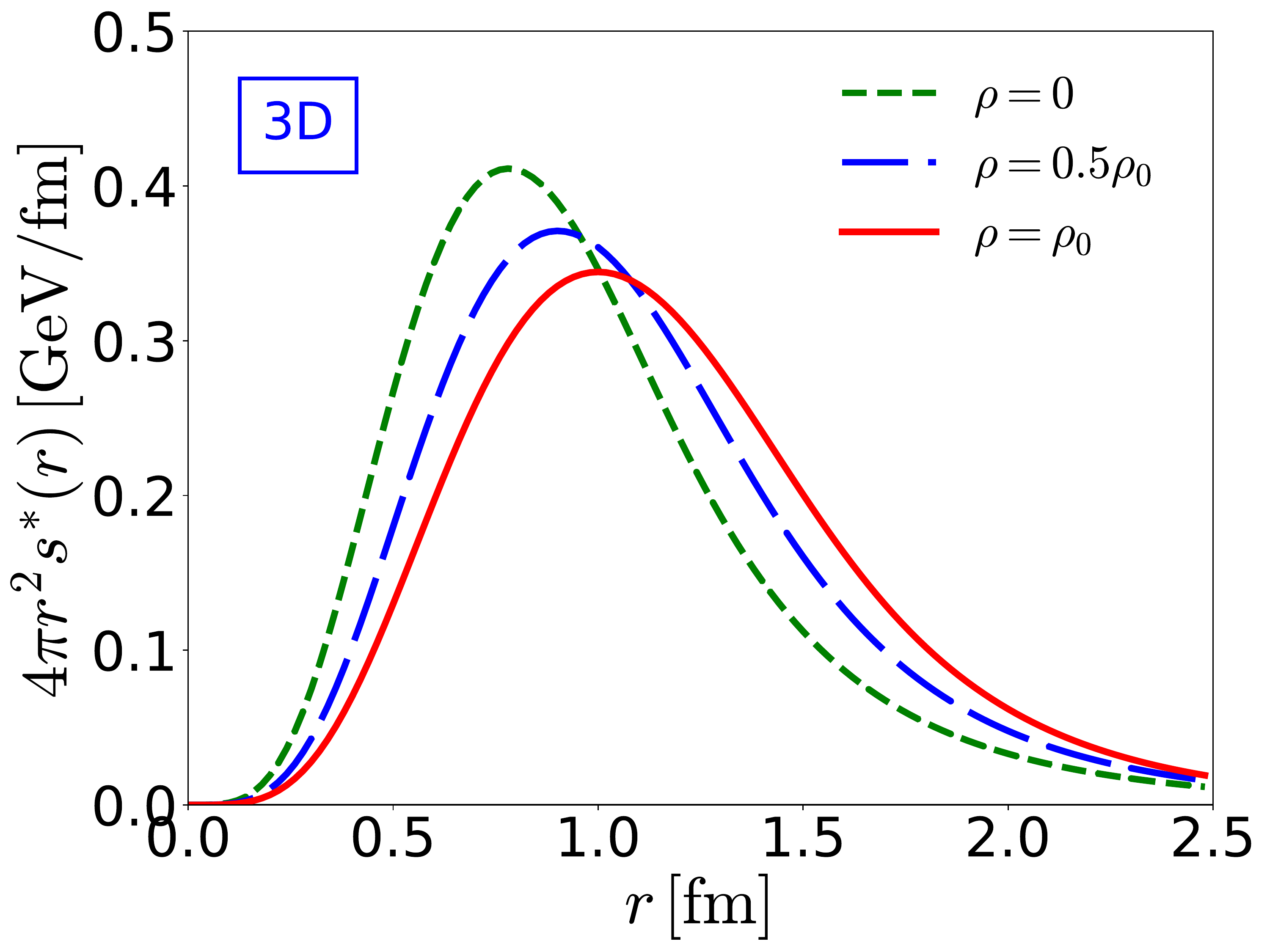}
\hspace{0.5cm}
\includegraphics[scale=0.27]{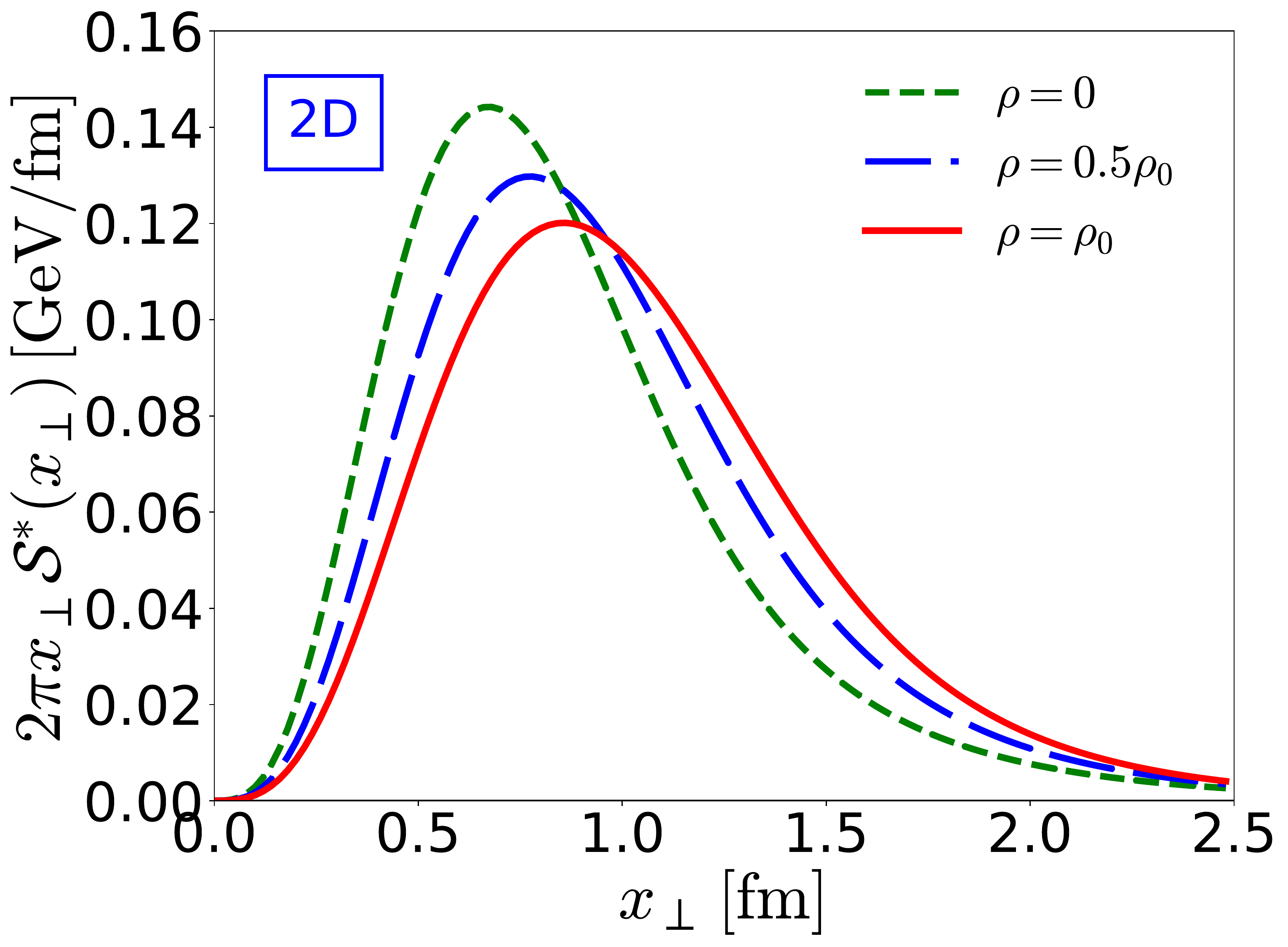}
\end{center}
\caption{
3D and 2D pressure distributions of the nucleon 
as functions of the nuclear density $\rho$, multiplied by $4\pi r^2$
and $2\pi x_\perp$, respectively. Notations are the same as in
Fig.~\ref{fig:5}. 
}
\label{fig:7}
\end{figure*}
In Fig.~\ref{fig:7}, we draw the 3D and 2D shear-force distributions of
the nucleon as functions of the nuclear density $\rho$, multiplied by
$4\pi r^2$ and $2\pi x_\perp$, respectively. Being similar to the case
of the pressure distributions, both the 3D and 2D shear-force
distributions get broadened as $\rho$ increases. Considering the
dependence of both the pressure and shear-force distributions on the
nuclear density, we can easily understand the reason why the 3D and 2D 
mechanical radii become larger in nuclear matter than in free
space. The 3D and 2D mechanical radii are defined in
Eq.~\eqref{eq:mech2d} in terms of the pressure and shear-force
distributions. Since they get more extended in nuclear matter, the
mechanical radii become larger as $\rho$ increases, as shown in
Table~\ref{tab:1}. Its physical implication is crucial because the
mechanical radius of the nucleon exhibits the physical size of the
nucleon. The results for the mechanical radii indicate that the
nucleon swells in nuclear matter. It also implies that the local
stability condition in Eq.~\eqref{eq:loc_stab} is satisfied. Its
positivity is indeed kept intact also in nuclear matter. 

It is of great interest to examine the ordering of the magnitudes of
the nucleon radii as done in Ref.~\cite{Kim:2021jjf}. Table~\ref{tab:1}
displays the following ordering for the 3D and 2D radii:
\begin{align}
\langle x_\perp^2\rangle_{\mathcal{E}} &< \langle
  x_\perp^2\rangle_{\mathrm{mech}} < \langle
  x_\perp^2\rangle_{J}\;\; (\mbox{2D radii}) ,\cr
\langle r^2\rangle_{\mathcal{E}} &< \langle
  r^2\rangle_{\mathrm{mech}} < \langle
  r^2\rangle_{J}\;\;\;\;\, (\mbox{3D radii}).  
\end{align}
While the ordering of the 2D radii is in agreement with that in the
chiral quark-soliton model~\cite{Kim:2021jjf}, the 3D
radii are ordered differently. We have the same orderings in nuclear
matter. 

\subsection{3D and 2D force fields inside the nucleon}

\begin{figure*}[htp]
\begin{center}
\includegraphics[scale=0.26]{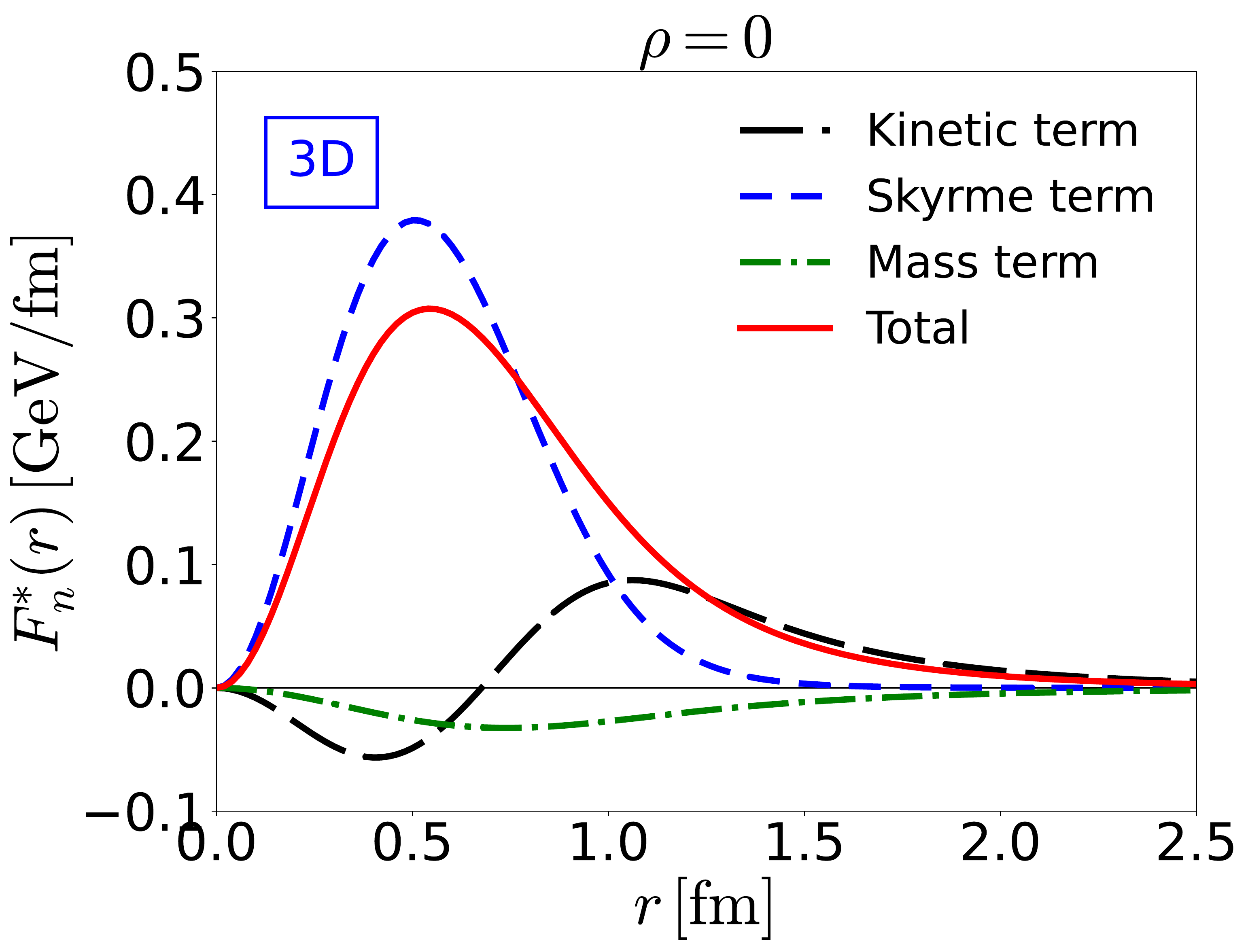}
\hspace{0.5cm}
\includegraphics[scale=0.26]{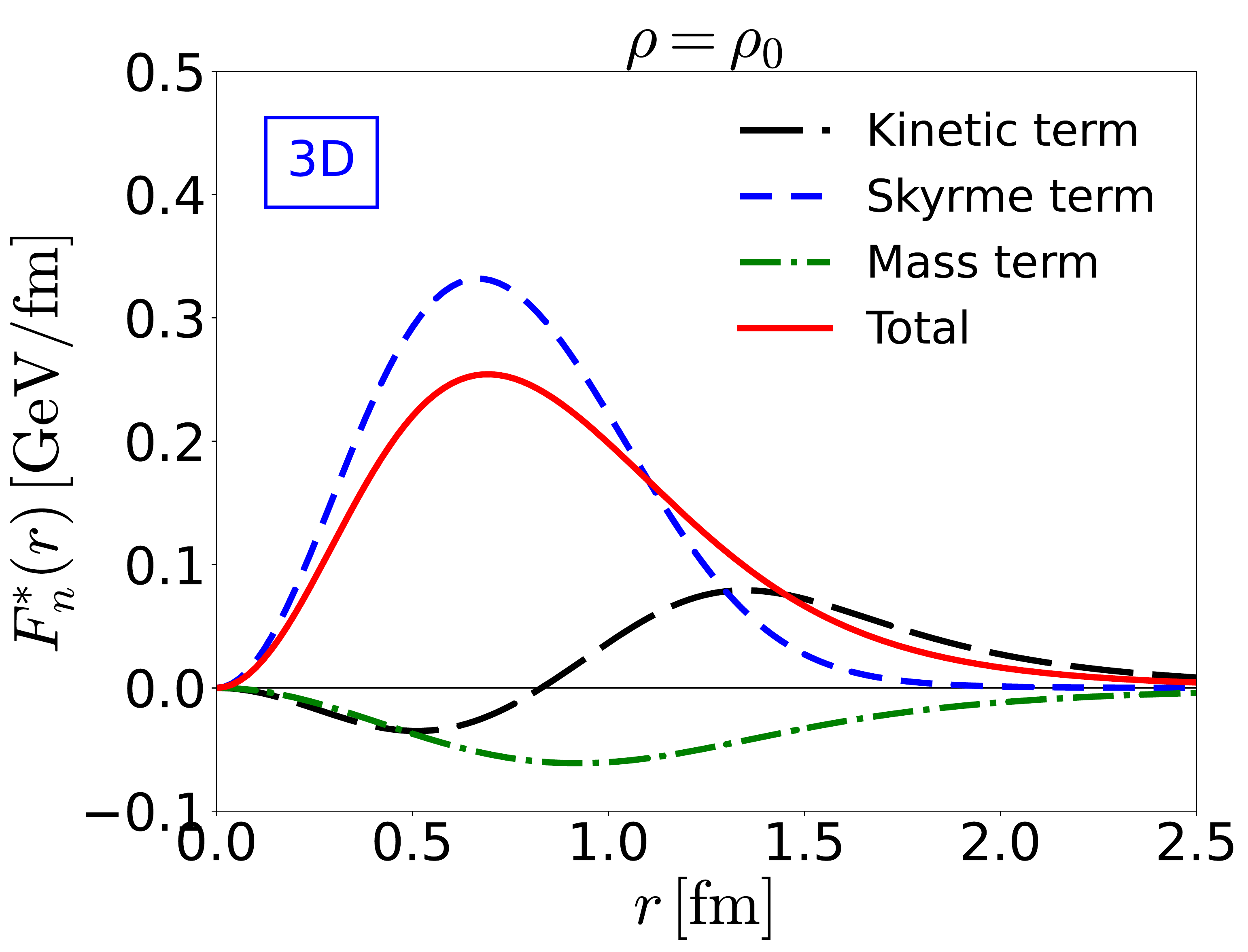}
\includegraphics[scale=0.26]{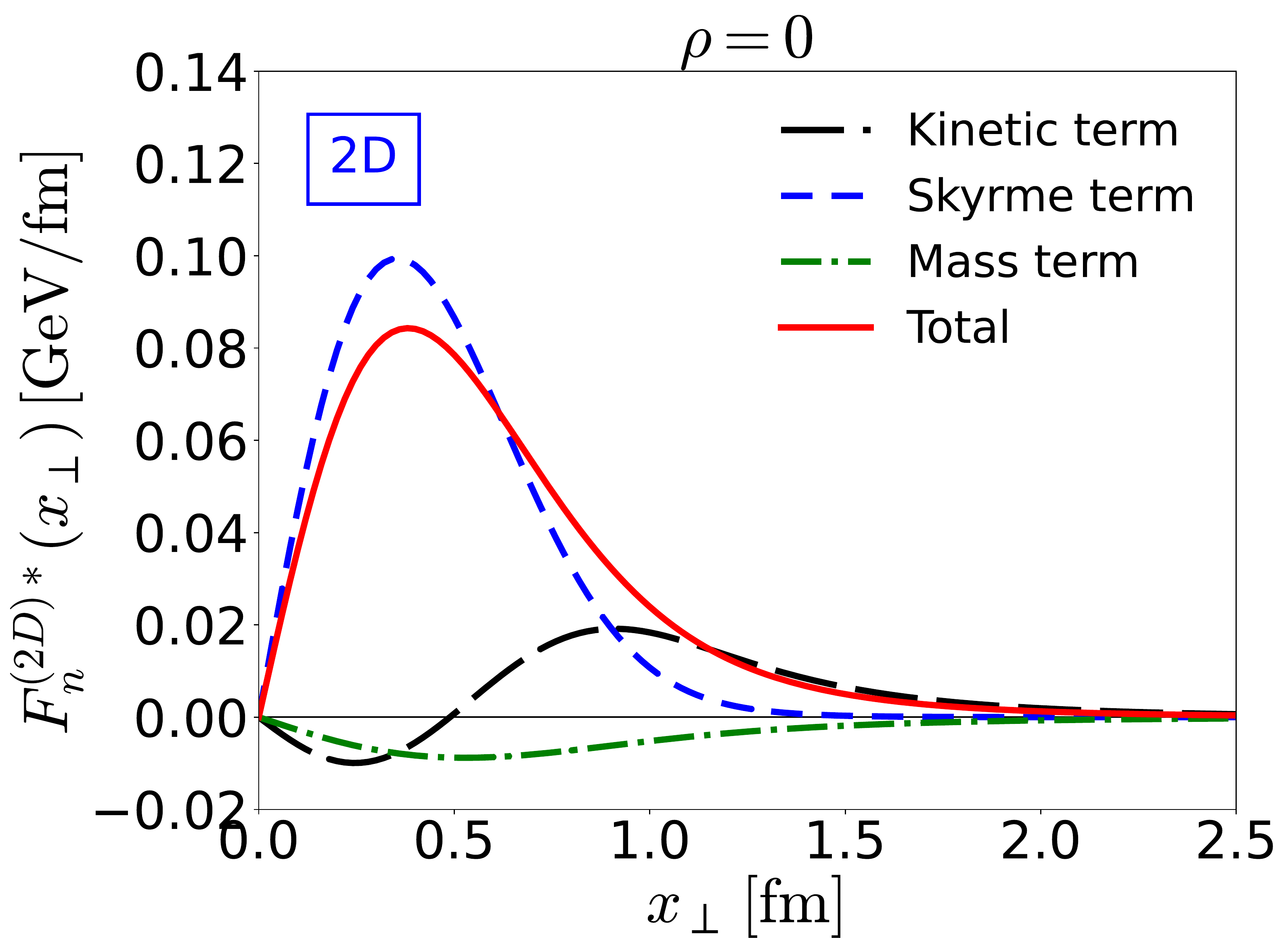}
\hspace{0.5cm}
\includegraphics[scale=0.26]{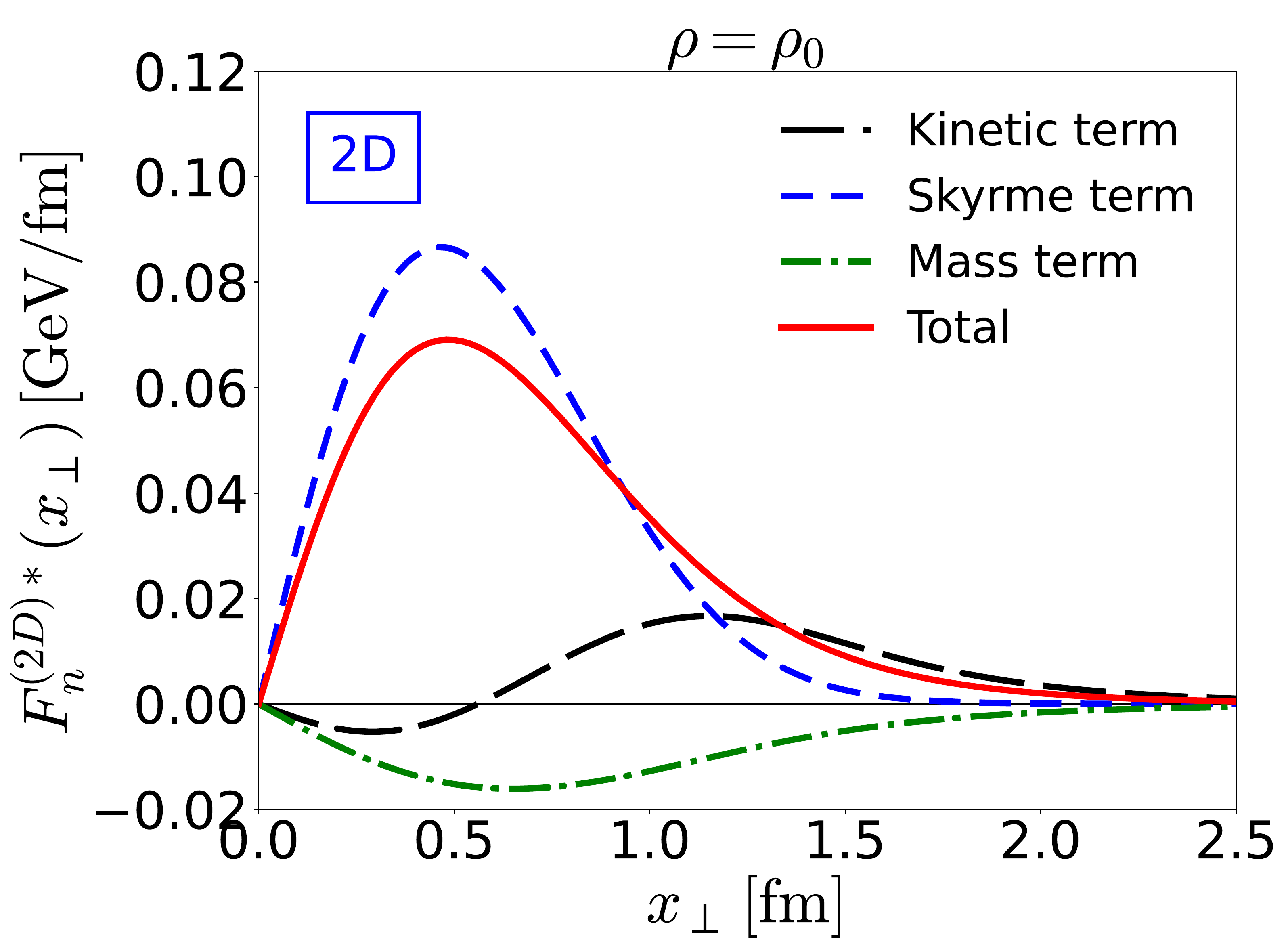}
\caption{3D and 2D normal force fields inside the nucleon.
The upper-left (-right) panel depicts the 3D normal force field
at $\rho=0$ ($\rho=\rho_{0}$). The lower-left (-right) panel draws the 2D
  force normal force fields at $\rho=0$ ($\rho=\rho_{0}$). Notations
  are the same as in Fig.~\ref{fig:2}. 
}
\label{fig:8}
\end{center}
\end{figure*}

\begin{figure*}[htp]
\begin{center}
\includegraphics[scale=0.26]{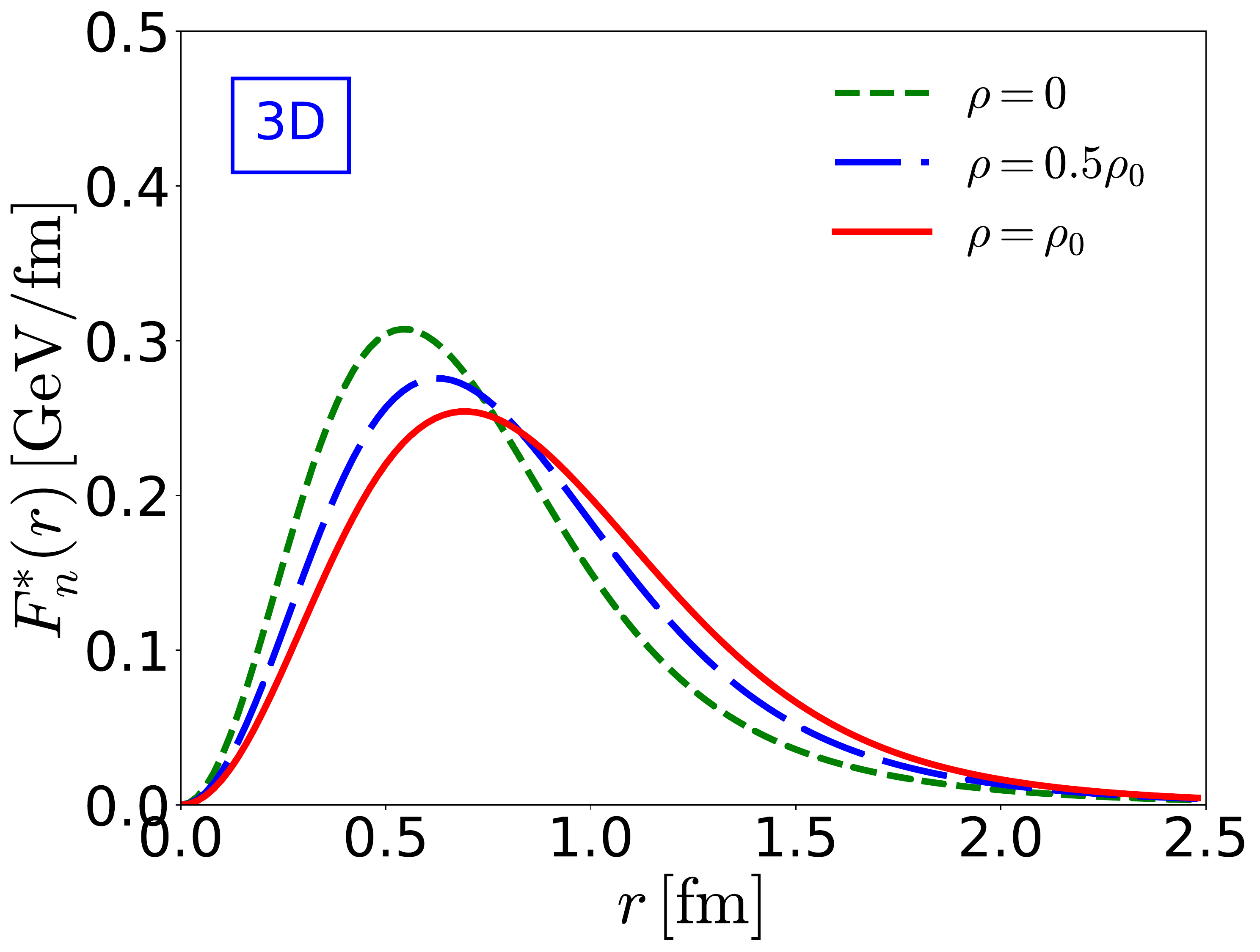}
\hspace{0.5cm}
\includegraphics[scale=0.26]{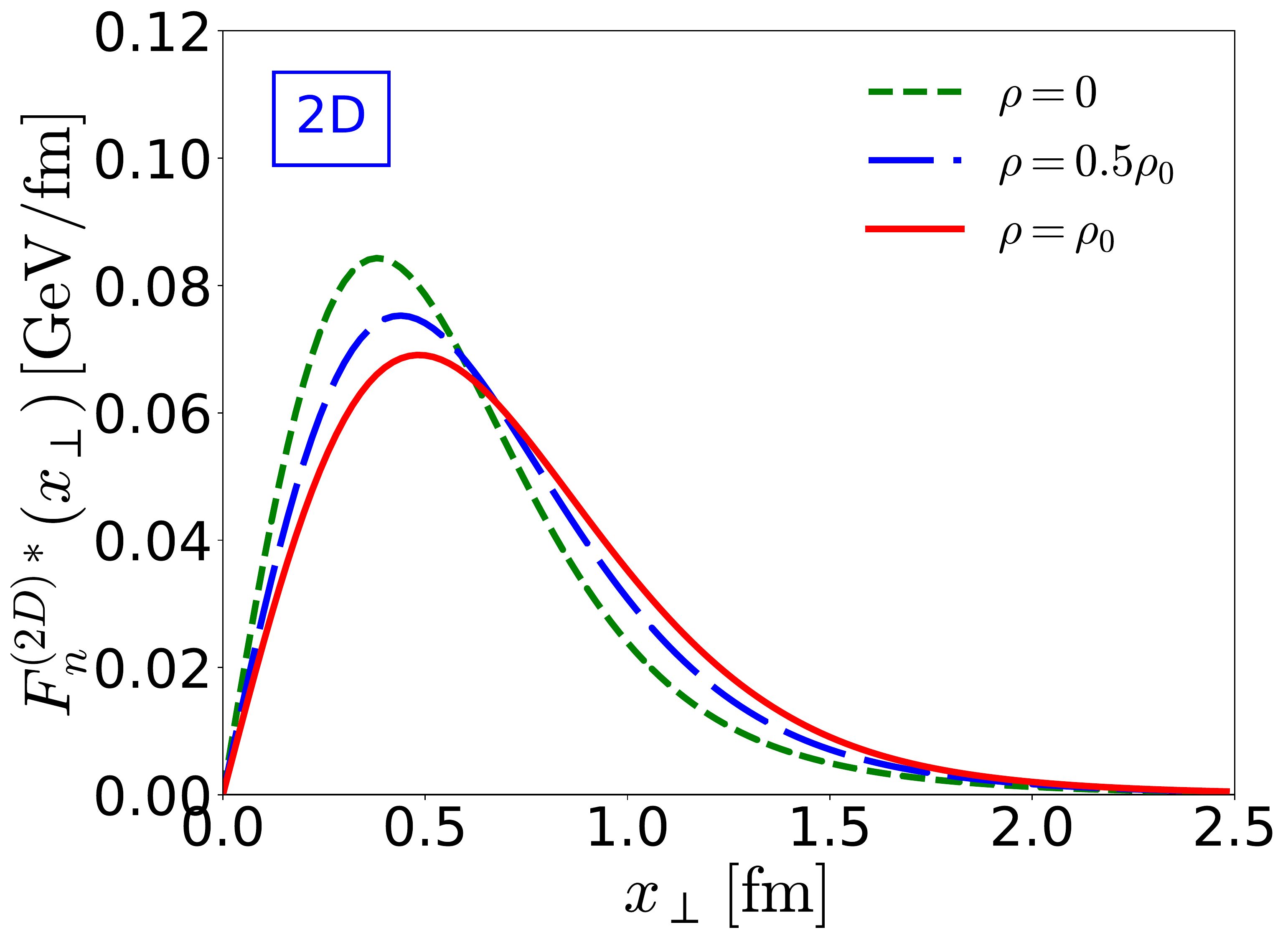}
\caption{3D and 2D normal force fields inside the nucleon 
as functions of the nuclear density $\rho$. The short-dashed,
long-dashed, and solid curves represent the normal force fields
with $\rho=0,\,0.5\rho_0, \rho_0$, respectively. $\rho_0$ denotes the
normal nuclear matter density.  
}
\label{fig:9}
\end{center}
\end{figure*}

\begin{figure*}[htp]
\begin{center}
\includegraphics[scale=0.26]{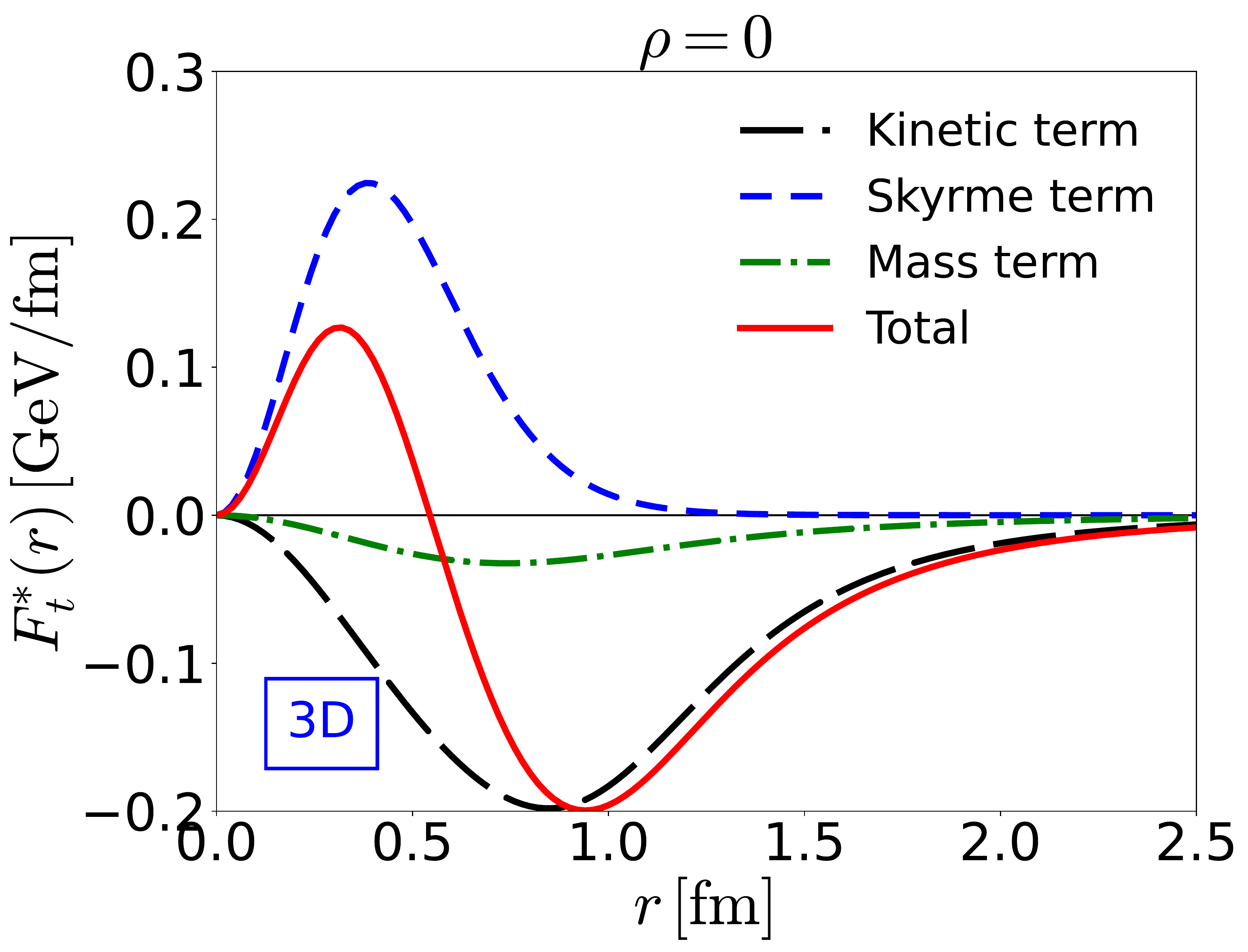}
\hspace{0.5cm}
\includegraphics[scale=0.26]{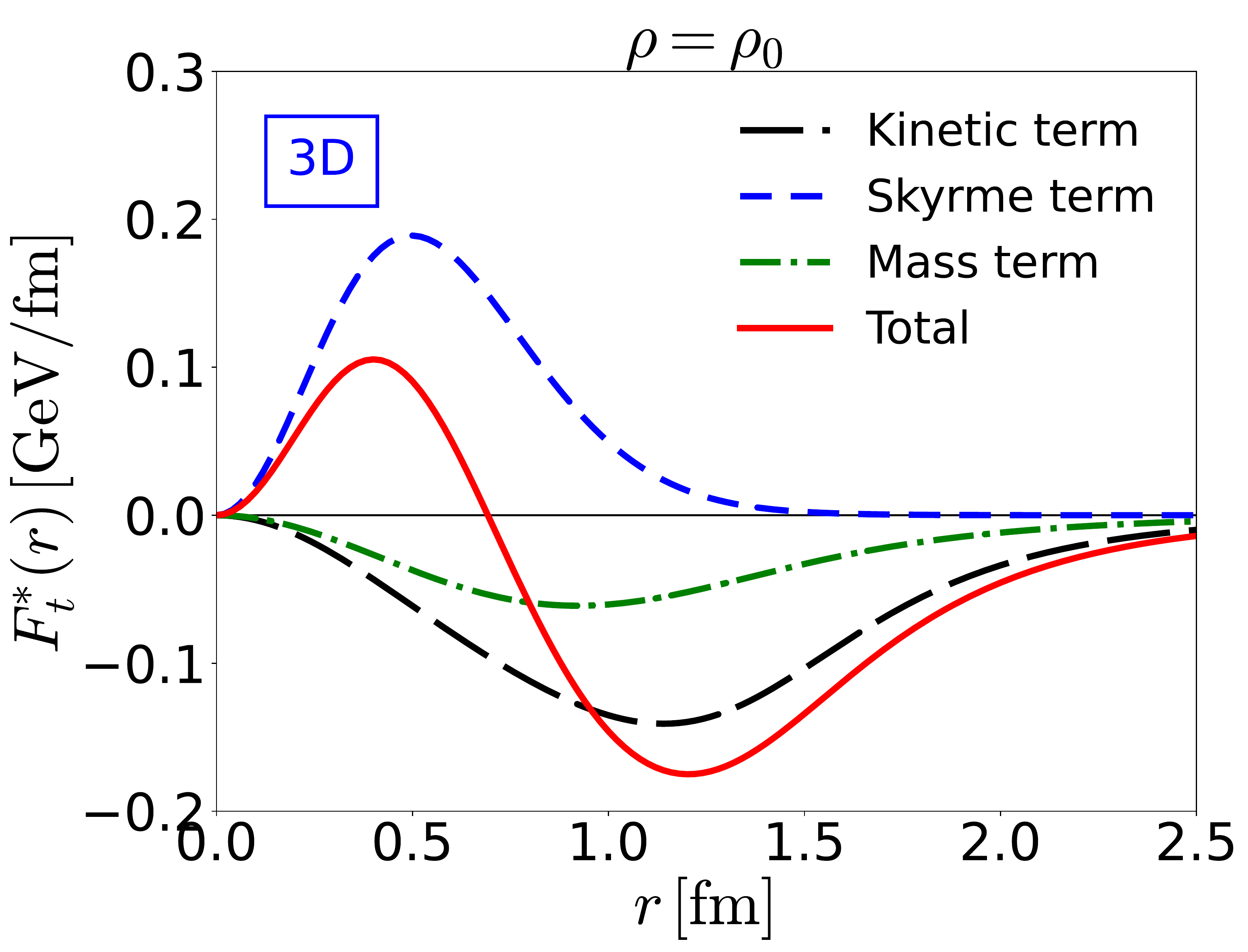}
\includegraphics[scale=0.26]{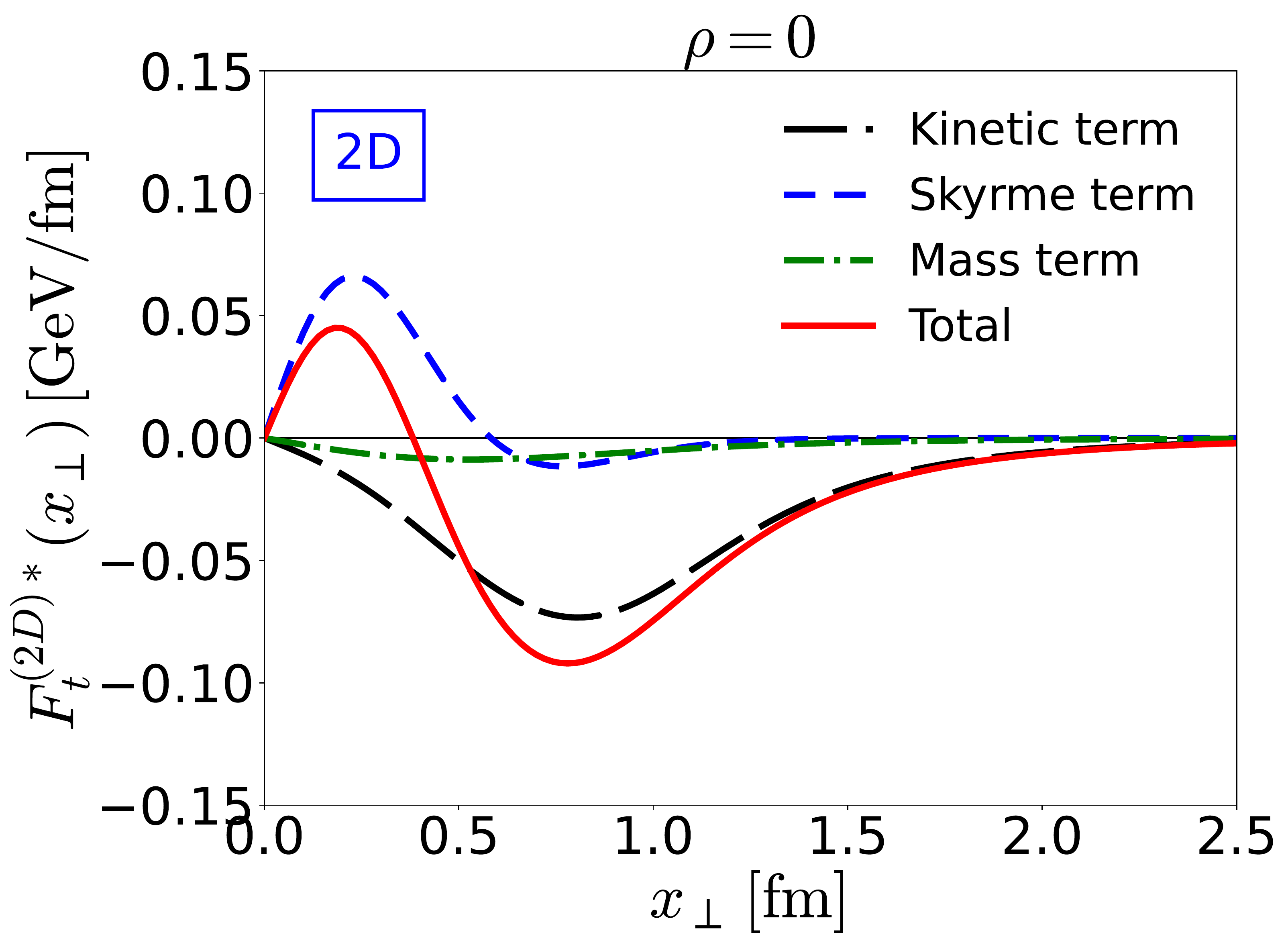}
\hspace{0.5cm}
\includegraphics[scale=0.26]{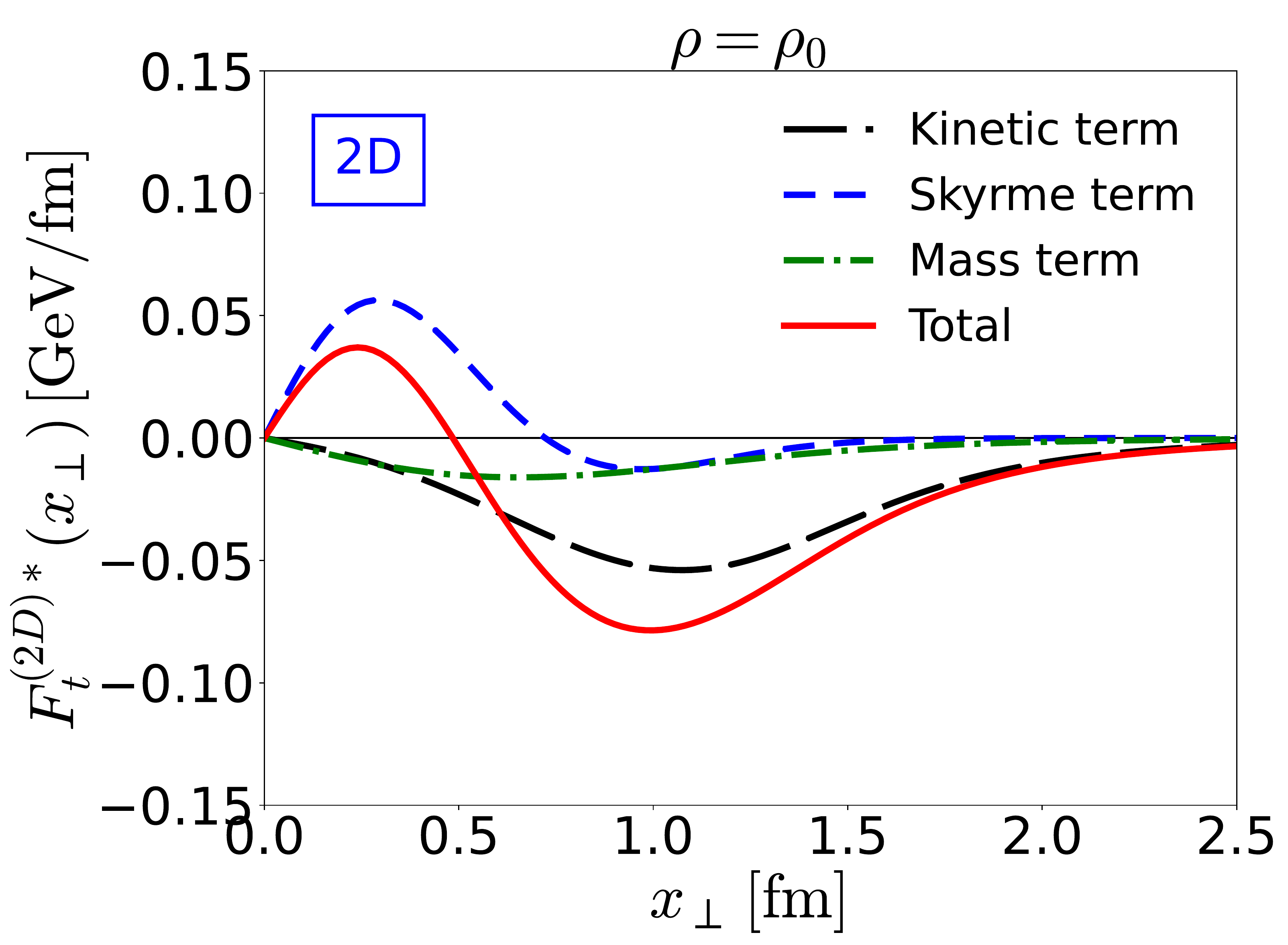}
\caption{3D and 2D tangential force fields inside the nucleon.
The upper-left (-right) panel depicts the 3D normal tangential field
at $\rho=0$ ($\rho=\rho_{0}$). The lower-left (-right) panel draws the 2D
  force tangential force fields at $\rho=0$ ($\rho=\rho_{0}$). Notations
  are the same as in Fig.~\ref{fig:2}. 
}
\label{fig:10}
\end{center}
\end{figure*}

\begin{figure*}[htp]
\begin{center}
\includegraphics[scale=0.26]{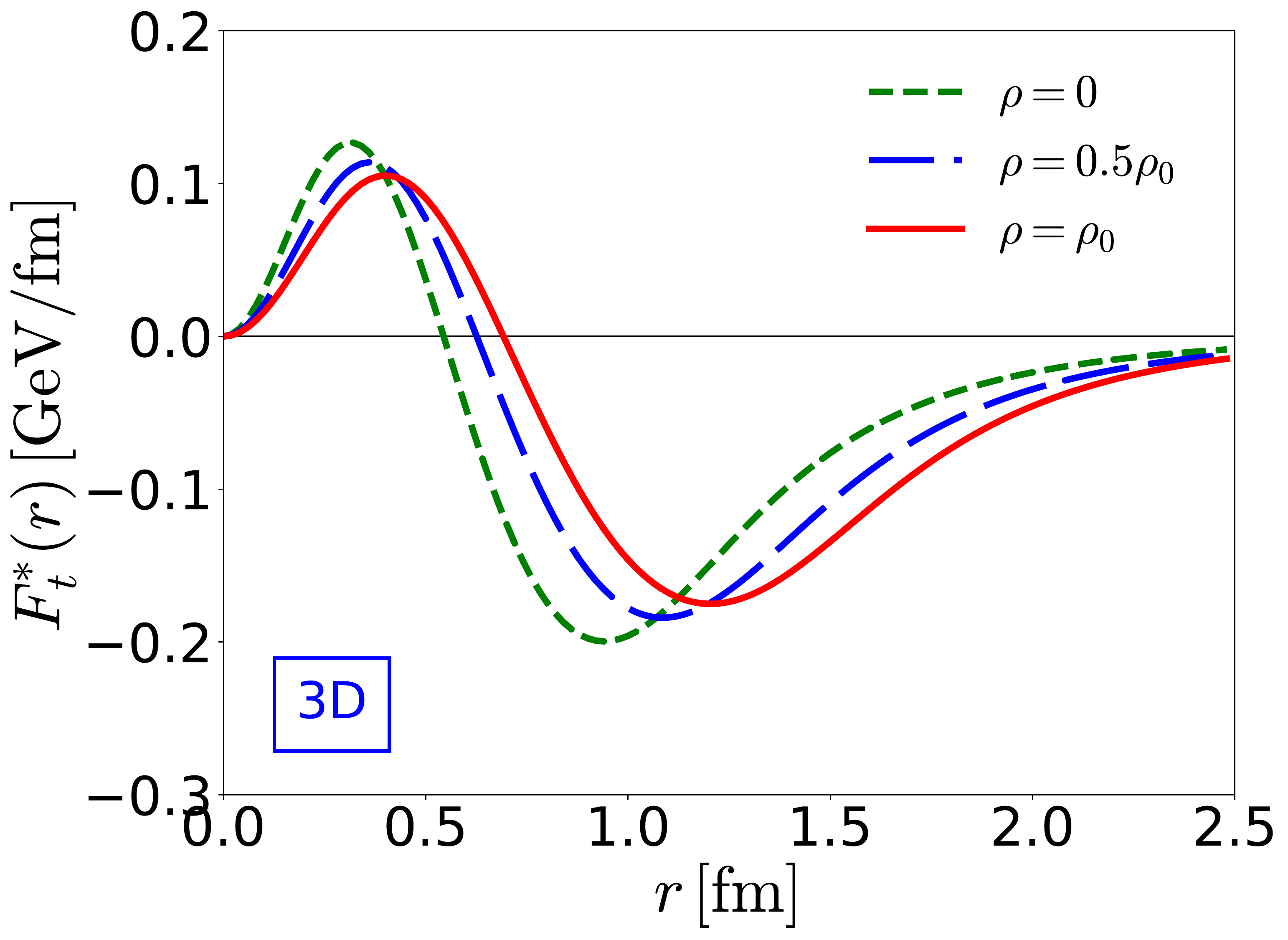}
\hspace{0.5cm}
\includegraphics[scale=0.26]{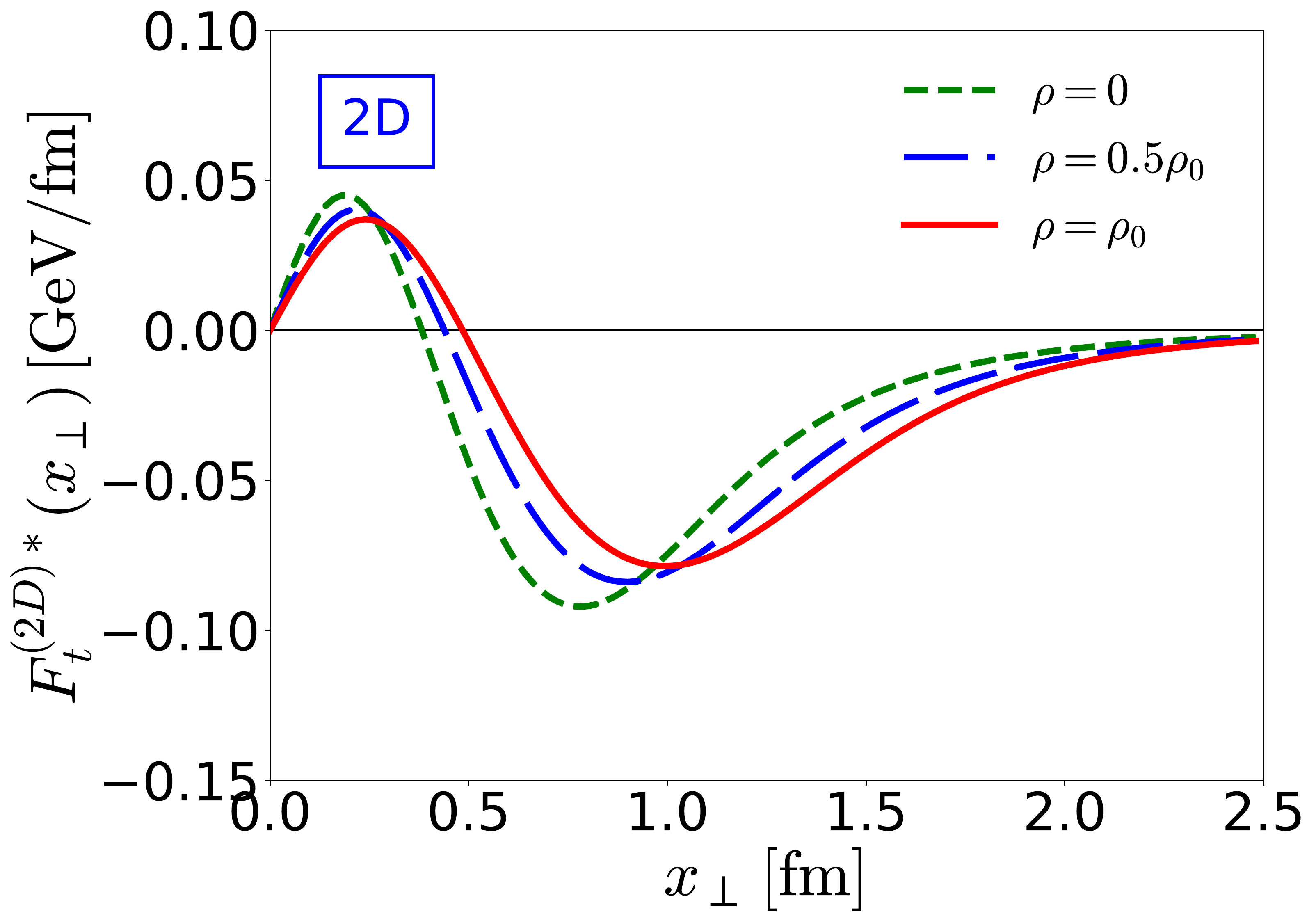}
\caption{3D and 2D tangential force fields inside the nucleon 
as functions of the nuclear density $\rho$. The short-dashed,
long-dashed, and solid curves represent the tangential force fields 
with $\rho=0,\,0.5\rho_0, \rho_0$, respectively. $\rho_0$ denotes the
normal nuclear matter density.  
}
\label{fig:11}
\end{center}
\end{figure*}
The strong force fields inside a nucleon reveal how the nucleon
acquires stability microscopically. As derived in
Eq.~\eqref{eq:2Dforce}, the normal and tangential force fields inside
the nucleon are nothing but the stability conditions given in
Eqs.~\eqref{eq:loc_stab} and~\eqref{eq:stab}: the 3D and 2D
normal force fields should be positive over the whole ranges, and the
tangential force fields should have at least one nodal point to secure
the stability condition. Figure~\ref{fig:8} illustrates how the normal
force field satisfies the local stability 
condition. As shown in the upper-left panel of Fig.~\ref{fig:8}, the
positivity of $F_n(r)$ is achieved in a nontrivial way. The Skyrme
term plays an essential role in making $F_n(r)$ satisfy the local
stability condition. Interestingly, the stable
topological soliton arises when the Skyrme term is included. It is
noticeable that this characteristic of the Skyrme 
model is reflected in the local stability condition. On the other
hand, the kinetic term negatively contributes to the core part
but positively to the outer shell. The mass term always provides
a negative contribution. The lower-left panel of Fig.~\ref{fig:8}
shows that the Abel transformation does not change the main feature of 
each contribution. As shown in the right panels of Fig.~\ref{fig:8},
both the 3D and 2D normal force fields become broadened in nuclear
matter. Figure~\ref{fig:9} exhibits how the normal force fields
undergo this broadening as $\rho$ increases. 

The upper-left panel of Fig.~\ref{fig:10} reminds us of the pressure
distributions in Fig.~\ref{fig:4}. The tangential force
field is just the integrand of the global stability condition or the
von Laue condition as shown in Eq.~\eqref{eq:stab}. Thus, the role of
each contribution is the same as in the case of the pressure
distributions as revealed in the upper-left panel of
Fig.~\ref{fig:10}. It is kept unmarred after the Abel transformation, 
as shown in the lower-left panel of Fig.~\ref{fig:10}. 
Figure~\ref{fig:11} demonstrates convincingly that 
the tangential force fields also get broadened as $\rho$ increases. 
Note that the sign of the 2D tangential force field turns negative at
around $x_\perp\approx 0.39$ fm in free space whereas at around
$x_\perp\approx 0.48$ fm at $\rho=\rho_{0}$. 

\begin{figure*}[htp]
\begin{center}
\includegraphics[scale=0.8]{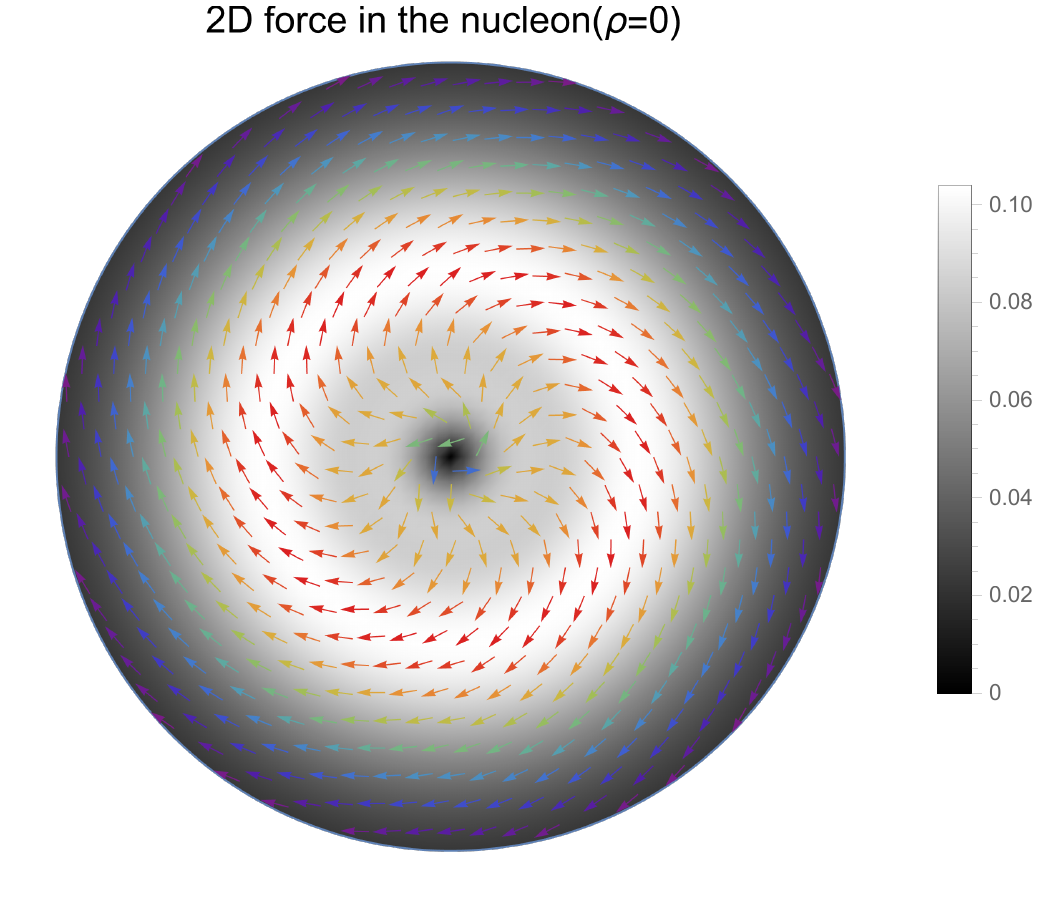}
\includegraphics[scale=0.8]{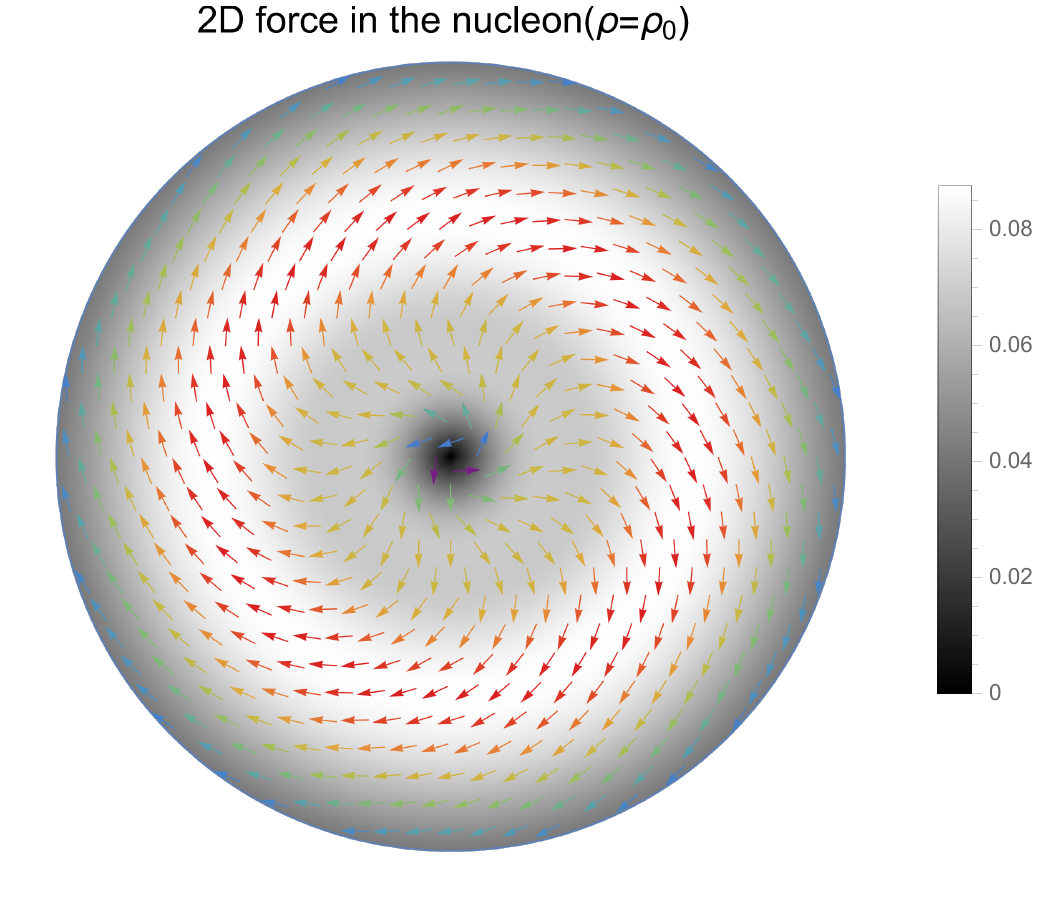}
\end{center}
\caption{Visualization of the 2D strong force field inside the
  nucleon. The left panel illustrates that in free space, whereas the
  right panel displays that in nuclear matter at the normal nuclear
  matter density.}
\label{fig:12}
\end{figure*}
The total strong force field in free space (nuclear matter) is
visualized in the left (right) panel of Fig.~\ref{fig:12}. The core
part of the strong force field is dominated by the normal force field, 
whereas the outer shell is governed by the tangential one. This
behavior of the strong force field ensures that the nucleon acquires
stability. In Ref.~\cite{Alharazin:2020yjv}, it was shown that 
 $F_t(x_\perp)$ comes into leading play at large distances in a 
model-independent way if one uses the $r$ dependence 
of the pressure and shear-force distributions. Last but not least, we 
want to mention that the total internal force fields include also
 contributions from the opposite direction to the force fields drawn 
 in Fig.~\ref{fig:12}. So, the nucleon is kept to be static. 

\section{Summary and outlook}

In the present work, we aimed at investigating the gravitational form
factors of the nucleon and the corresponding mechanical distributions
in nuclear matter within the framework of the in-medium modified Skyrme
model. We first computed the three different gravitation form factors
of the nucleon: the mass, angular momentum, and $D$-term form factors
with the nuclear density varied from $\rho=0$ to $\rho=\rho_0$,
where $\rho_0$ denotes the normal nuclear matter density. As the
momentum transfer increases, the mass and angular momentum form
factors fall off faster than the case in free space with $\rho$
increased. Since they are normalized to $A^{*}(0)=1$ and 
$J^{*}(0)=1/2$, the magnitudes of these two form factors at $t=0$ are not
changed. On the other hand, the absolute magnitude of the $D$-term
form factor is enhanced in nuclear matter and drops off more rapidly
than in free space. We then examined the mass distribution of the
nucleon in nuclear matter. Each contribution of the kinetic term, the
Skyrme term, and the mass term becomes broader as $\rho$
increases. This feature is kept unchanged by the Abel transformation
except for the mass term. The contribution of the mass term turns
negative in the two-dimensional transverse plane on the light
cone. The angular momentum distributions show similar behaviors 
as the nuclear density is changed. 

The pressure and shear-force distributions play crucial roles in
understanding the stability of the nucleon. The pressure distribution
has at least one nodal point so that the von Laue condition is
satisfied. The kinetic term and mass term of the pion provide
attraction whereas the Skyrme term yields repulsion. As the nuclear
density increases, both the 3D and 2D pressure distributions becomes
broader. The shear-force distributions exhibit the density dependence,
being similar to the pressure distributions. The Abel transformations
do not change the general features of the pressure and shear-force
distributions. The local and global stability conditions are satisfied
also in nuclear matter. The mass, angular momentum, and mechanical radii
of the nucleon become larger in nuclear matter than in free space,
which indicates that the nucleon swells in nuclear matter. 

Finally, we scrutinized the strong normal and tangential force fields,
which are deeply related to the stability conditions. The normal force
fields are positive over the whole space and transverse plane. The 2D
strong force fields keep all important features of the 3D force
fields, which means that the Abel transformation does not change the
characteristics of the force fields. In conclusion, the nucleon
acquires stability also in nuclear matter in a certain range of the
nuclear matter density. 

\section*{Acknowledgments}
The work was supported by the Deutscher Akademischer
Austauschdienst(DAAD) doctoral scholarship (J.-Y.K) and Basic Science
Research Program through the National Research Foundation of Korea
funded by the Korean government (Ministry of Education, Science and
Technology, MEST), Grant-No. 2020R1F1A1067876 (U. Y.) and
2021R1A2C2093368, 2018R1A5A1025563 (H.-Ch.K.). 


\end{document}